\documentclass[a4paper,fleqn]{cas-sc}

\usepackage{graphicx}
\usepackage{amsmath}
\usepackage{amssymb}
\usepackage{bm}

\numberwithin{equation}{section}

\newcommand{\red}[1]{{\color{black}#1}}

\newcommand{\Tr}{\operatorname{Tr}}

\begin{document}

\title{Non-renormalization of the Hall viscosity of integer and Jain fractional quantum Hall phases by Coulomb interactions}

\author[1]{M. Selch}[orcid=0009-0009-5715-7280]
\affiliation[1]{organization={Physics Department, Ariel University},
               city={Ariel},
               postcode={40700},
               country={Israel}}

\shorttitle{}
\shortauthors{M. Selch}

\begin{abstract}
We proof the non-renormalization of the Hall viscosity by Coulomb interactions for integer and Jain fractional quantum Hall states in the presence of Galilean invariance building on previous results obtained for the Hall conductivity. We employ Wigner-Weyl calculus in order to represent the Hall viscosity in terms of a novel Green function expression supplemented by covariant derivatives. The Green function based expression is derived within the free field theory of electrons and explicitly calculated for this case as well as in the mean field approximation of the composite fermion theory Jain states employing the model by Lopez and Fradkin. The topological orbital spin of composite fermions distinguishes their mean field treatment from that of electrons resulting in an additional topological contribution. We furthermore proof the topological invariance of the newly introduced Green function expression for the Hall viscosity for the leading contribution in powers of the filling factor with and without Galilean boost symmetry. The latter condition is known to already imply a Hall viscosity per emergent quasiparticle number density quantized in units of one half times the average quasiparticle orbital spin or one quarter times the Wen-Zee shift. The Wen-Zee shift features a contribution from the composite fermion topological orbital spin relative to that of electrons which we highlight within our formalism and connect to Haldane's geometric Hall viscosity framework. 
\end{abstract}


\maketitle
\section{Introduction}
\label{Introduction}

The quantum Hall effect (QHE) stands as one of the most profound manifestations of topological order in condensed matter physics. Initially it was observed through quantized plateaus in the Hall resistivity of two-dimensional electron systems subjected to strong magnetic fields in both its integer form \cite{klitzing1980quantized} as well as its fractional form \cite{tsui1982two}. The phenomenon has since inspired a broader understanding of topological phases of matter. These phases are characterized by global, quantized entities insensitive to local perturbations or microscopic details. Among these is the more recent Hall viscosity, a non-dissipative response to shear deformations, which has emerged as a geometric response function that encodes topological information of the underlying quantum state complementary to the quantized Hall conductivity \cite{avron1995viscosity,avron1998odd}. Both topological responses are being used to tell topological phases of matter apart.\par
Unlike conventional viscosity, which quantifies energy dissipation in fluids, Hall viscosity arises in systems where time-reversal and parity symmetry are broken and is intimately tied to the Berry curvature of the many-body wavefunctions in momentum space \cite{read2009non}. It represents a response coefficient that links strain rate to a transverse stress and is uniquely non-dissipative. Its existence is particularly salient in gapped quantum fluids such as integer and fractional quantum Hall (FQH) states where it has been studied initially.\par
Despite compelling theoretical predictions, direct experimental access to Hall viscosity in two dimensional electron systems remained elusive. The key challenge lies in the fact that Hall viscosity does not couple to conventional electromagnetic probes, as it pertains to internal stress in the form of strain rather than charge transport. However, it has been found that Hall viscosity subtly influences electrochemical potentials locally on finite spatial distance scales, leading to measurable corrections in electrostatic profiles and current distributions \cite{hoyos2012hall,hoyos2014hall}.
Further detailed theoretical studies employing kinetic theory and hydrodynamics for electron flow  \cite{alekseev2016negative,sherafati2016hall, scaffidi2017hydrodynamic,delacretaz2017transport,pellegrino2017nonlocal} paved the way for detailed experimental studies. The line of reasoning culminated in the first experimental detection of Hall viscosity in 2019, as reported in \cite{berdyugin2019measuring}. In this study, the authors examined current flow in ultra-clean graphene in a magnetic field under conditions where electron-electron interactions dominate, and hydrodynamic behavior emerges. By carefully measuring nonlocal resistance and variations in the electrochemical potential over finite distances near current injectors, they were able to extract the Hall component of the viscous response. The results provided the first empirical confirmation of a quantity previously considered primarily theoretical, opening new avenues for probing geometric and topological properties of quantum fluids.\par
Additional channels for Hall viscosity measurements have been proposed like chiral vector field induction by acoustic and optical phonons in Dirac materials \cite{cortijo2015hall,heidari2019hall}. Hall viscosity shows up via a shift in the dispersion relation of optical phonons at zero momentum. Hall viscosity in superfluid helium and phonon systems may induce the acoustic Faraday effect \cite{tuegel2017hall,ye2021phonon} and further contribute to the thermal Hall conductivity of phonons \cite{ye2021phonon}. Moreover, a nonzero Hall viscosity induces elliptical polarization of sound waves in the superfluid phases of helium \cite{furusawa2021hall,fujii2018low}. Coupling a Hall fluid with non-vanishing Hall viscosity to photons may induce non-trivial optical topological phases \cite{mechelen2019viscous,mechelen2021optical}. This may allow to probe Hall viscosity via surface magneto-optic Kerr effect spectroscopy.\par
A striking feature of Hall viscosity is its potential topological invariance. In certain regimes, particularly for incompressible quantum Hall states, the Hall viscosity to density ratio is found to be quantized and proportional to the so-called Wen-Zee ``shift'', a topological quantum number appearing in the relation between the total magnetic flux and the number of particles on a curved manifold \cite{read2011hall}. The Hall viscosity is furthermore intimately related to the Hall conductivity for Galilean invariant fluids as established by Ward identities \cite{bradlyn2012kubo} and effective field theory methods \cite{hoyos2012hall}. This observation raises profound questions such as under which conditions is the Hall viscosity a topological invariant relative to Hall conductivity. In a broader context topological invariants reveal intimate relations between solid state systems and high energy physics \cite{zubkov2012momentum} as exemplified by fermionic superfluids on the one hand \cite{volovik2013nambu} and QCD and the standard model of particle physics on the other hand \cite{zubkov2018momentum,volovik2017standard}.\par 
This paper explores the question of the topological invariance of the Hall viscosity further by analyzing its robustness in the presence of Coulomb interactions for integer and Jain quantum Hall states. Our discussion relies on Galilean invariance even though the quantization of the Hall viscosity over particle number density ratio has been found without insisting on Galilean boost symmetry. Using both non-relativistic free field theory and effective composite fermion field theory together with Wigner-Weyl calculus we derive a novel Green function expression for the Hall viscosity which is supplemented by covariant derivatives. We examine the range of validity of the Hall viscosity Green function expression and find it to be topologically invariant at leading order in the Landau level filling factor. 
Early field theory treatments of quantum Hall physics may be found in \cite{jain1989composite,jain1990theory,Zhang1989effective}. We appeal to the formulation by Lopez and Fradkin \cite{Fradkin1991chern} together with its relativistic extension as discussed in \cite{selch2025hall} in distinction to the recently introduced particle-hole symmetric, lowest Landau level Dirac composite fermion theory \cite{son2015composite,son2018dirac,nguyen2018particle}. While the topological invariance of the Hall conductivity has been known for a long time \cite{TKNN2}, the perturbative proof of (anomalous) Hall conductivity non-renormalization \cite{ZZ2019_0,ZZ2021} as well as a more in depth analysis of the topological robustness in the presence of inhomogeneous magnetic fields \cite{ZW2019,ZZ2019_1}, and its breakdown at finite temperatures and real time variations \cite{Banerjee2021}, framed within Wigner-Weyl formulation, have been performed only recently.\par
Our studies extend the topological robustness analysis within Wigner-Weyl calculus Green function representations as performed for the Hall conductivity to Hall viscosity in a novel way. It is noteworthy that Green function topological expressions representing quantized response functions for the entire periodic table of non-interacting topological insulators have been found \cite{essin2011bulk}. The standard Green function topological expression for the Hall conductivity seems to work only for non-interacting theories where it represents the first Chern number. In the case of integer quantum Hall phases it also coincides with the so-called Wen-Zee shift which is a geometric quantum number and happens to parametrize Hall viscosity over particle number density ratio quantization. It appears that the mentioned Green function topological expression for fractional quantum Hall phases, which are strongly correlated, follows the Wen-Zee shift rather than the quantized Hall response, though not for all Jain states \cite{gurarie2013topological,gurarie2014hall}. The Wen-Zee shift equals twice the so-called average orbital spin. We initiate the study of a new candidate Green function expression capturing the topology of the Hall viscosity. As the composite fermion theory enables a restriction to integer quantum Hall physics even for fractional quantum Hall states, we avoid difficulties deriving from strong correlation effects appearing beyond mere topological effects. Relative to Haldane's geometric discussion of the Hall viscosity \cite{haldane2009hall,haldane2011geometrical,park2014guiding} we also capture the Landau orbit spin and guiding center spin contributions as constituents of the average orbital spin within the composite fermion formulation, but under stricter symmetry assumptions.\par
Further (topological) response coefficients beyond the Hall conductivity and the here studied Hall viscosity which have been discussed within Wigner-Weyl calculus are, e. g., the scale magnetic \cite{chernodub2017scale} and chiral separation \cite{zubkov2023effect} effects. We will make use of our recent results on topological invariance and perturbative non-renormalization of the Hall conductivity in \cite{selch2025hall} which will be particularly useful in applications to the Hall viscosity.\par
Our work is structured as follows. In Section~\ref{Weylsymbol}, we provide a review of essentials of the Wigner-Weyl calculus used throughout. Section~\ref{SectnonRel} introduces the Hall viscosity in general terms and then specifically in the context of quantum Hall systems as described by the field theory model of Lopez and Fradkin. We explicitly calculate the Hall viscosity at the mean field level and analyze the topological robustness of the Hall viscosity Green function expression in Galilean invariant systems breaking time reversal symmetry and parity in the non-interacting case. We then initiate the proof of non-renormalization under perturbative Coulomb interactions. We proceed using the results of our recent work \cite{selch2025hall} for the completion of the proof presented in Section~\ref{Sectnonrenormalization}. Our main findings will be reviewed in Section~\ref{SectDiscussion} with Section ~\ref{SectConclusions} containing our conclusions. Details of formalism and involved calculations are reserved for several appendices. 

\section{The Weyl-symbol of an operator}
\label{Weylsymbol}

We start out with the basic definitions within Wigner-Weyl calculus taken mostly from appendix B of \cite{selch2023hall}, as we will both heavily rely on it in the course of this work and its application is not so wide spread. At the same time it serves to explain our notational conventions. On the one hand it provides a bridge between the canoncial formulation of quantum mechanics in terms of operators and states on a Hilbert space and functions on phase space. It is an equivalent reformulation of the quantum physics on Hilbert spaces assuming Weyl-ordering of operators. A description by functions on phase space is familiar from classical physics and may be interpreted as getting deformed by quantum physics due to finite Planck constant $h$. In other words the quantum phase space formulation reduces to the classical one in the limit $h\to 0$. On the other hand the Wigner-Weyl transformation of an operator to a phase space function allows for an additional way of representing physics as compared to that in configuration or momentum space (which are related by Fourier transformation). It is suitable whenever a system is close to being homogeneous which we are assuming. The Wigner-Weyl calculus establishes a one-to-one correspondence between a quantum mechanical theory defined on a Hilbert space and its reformulation in terms of functions on phase space. An operator $\hat{\mathcal{O}}$ on Hilbert space, being a function of the position operator $\hat{x}$ and the momentum operator $\hat{p}$ in $D$ spacetime dimensions, is associated with a phase space function $O_W(x,p)$ of space coordinate $x$ and momentum coordinate $p$ by the definition
\begin{align}
\hat{\mathcal{O}}=\int \frac{d^Dk}{(2\pi )^D}\frac{d^Dp}{(2\pi )^D}d^Dyd^DxO_W(x,p)e^{i(k(x-\hat{x})+y(p-\hat{p}))}.
\end{align}
A wave function on Hilbert space is represented in bra-ket notation by $|\Psi\rangle$ with configuration space representation $\Psi (x)=\langle x|\Psi\rangle$. We would like to relate the phase space function $O_W(x,p)$ to the configuration space representation of $\hat{\mathcal{O}}$ denoted by $O(x_1,x_2)=\langle x_1|\hat{\mathcal{O}}|x_2\rangle$ as well as to its momentum space representation denoted by $\tilde{O}(p_1,p_2)=\langle p_1|\hat{\mathcal{O}}|p_2\rangle$. The configuration and momentum space representations are related by
\begin{align}
O(x_1,x_2)=\int\frac{d^Dp_1}{(2\pi )^{\frac{D}{2}}}\frac{d^Dp_2}{(2\pi )^{\frac{D}{2}}}e^{ip_1x_1}\tilde{O}(p_1,p_2)e^{-ip_2x_2}.
\end{align}
In order to achieve this we calculate $\langle z|\hat{\mathcal{O}}|\Psi\rangle$. Applying the identities
\begin{align}
\langle z|e^{ik\hat{x}}|\Psi\rangle =e^{ikz}\langle z|\Psi\rangle =e^{ikz}\Psi (z),\,\,\,\,\langle z|e^{-iy\hat{p}}|\Psi\rangle =e^{-y\partial_z}\langle z|\Psi\rangle =e^{-y\partial_z}\Psi (z) =\Psi (z-y)
\end{align}
as well as the Baker-Campbell-Hausdorff relation
\begin{align}
e^{-i(k\hat{x}+y\hat{p})}=e^{-ik\hat{x}}e^{-iy\hat{p}}e^{i\frac{ky}{2}}
\end{align}
we obtain
\begin{align}
\nonumber \langle z|\hat{\mathcal{O}}|\Psi\rangle &=\langle z|\int \frac{d^Dk}{(2\pi )^D}\frac{d^Dp}{(2\pi )^D}d^Dyd^DxO_W(x,p)e^{i(k(x-\hat{x})+y(p-\hat{p}))}|\Psi\rangle\\
\nonumber &=\int \frac{d^Dkd^Dp}{(2\pi )^{2D}}d^Dyd^DxO_W(x,p)e^{i(kx+yp+\frac{ky}{2})}e^{-ikz}\Psi (z-y)\\
\nonumber &=\int \frac{d^Dp}{(2\pi )^D}d^Dyd^DxO_W(x,p)\delta^D\Big(x+\frac{y}{2}-z\Big)e^{iyp}\Psi (z-y)\\
&=\frac{1}{(2\pi )^D}\int d^Dpd^DyO_W\Big(z-\frac{y}{2},p\Big)e^{ipy}\Psi (z-y).
\end{align}
A comparison with the configuration space relation $\langle z|\hat{O}|\Psi\rangle =\int d^DxO(z,x)\Psi (x)$ for $x=z-y$ implies
\begin{align}
O(x,y)=\frac{1}{(2\pi )^D}\int d^DpO_W\Big(\frac{x+y}{2},p\Big)e^{ip(x-y)}.
\label{toderivestarproduct}
\end{align}
An inversion may be performed by a change of variables $R=\frac{x+y}{2}$ and $r=x-y$ followed by a Fourier transformation with respect to $r$. It reads
\begin{align}
O_W(R,p)=\int d^DyO\Big(R+\frac{y}{2},R-\frac{y}{2}\Big)e^{-ipy}
\end{align}
The analogous relation for the momentum space representation may be obtained similarly and reads
\begin{align}
\tilde{O}(p,k)=\frac{1}{(2\pi )^D}\int d^DxO_W\Big(x,\frac{p+k}{2}\Big)e^{i(k-p)x}
\label{momentumww}
\end{align}
with inverse
\begin{align}
O_W(R,p)=\int \frac{d^Dk}{(2\pi )^D}\tilde{O}\Big(p+\frac{k}{2},p-\frac{k}{2}\Big)e^{ikR}.
\end{align}
The Weyl-symbol of the product $\hat{\mathcal{C}}=\hat{\mathcal{A}}\hat{\mathcal{B}}$ of two operators $\hat{\mathcal{A}}$ and $\hat{\mathcal{B}}$ on the Hilbert space may be represented as follows
\begin{align}
C_W(x,p)=A_W(x,p)\star B_W(x,p).
\label{weylsymbolproductofoperators}
\end{align}
The symbol $\star$ represents the Moyal star product
\begin{align}
\star = {\rm exp} \Big(\frac{i}{2} \left( \overleftarrow{(\partial_x)}^{\mu}\overrightarrow{(\partial_p)}_{\mu}-\overleftarrow{(\partial_p)}_{\mu}\overrightarrow{(\partial_x)}^{\mu}\right) \Big).
\end{align}
The form as written in Eq. (\ref{weylsymbolproductofoperators}) may be derived from Eqs. (\ref{toderivestarproduct}) and (\ref{momentumww}) and employing partial integration as follows
\begin{align}
\nonumber C_W(R,p)=&\int d^DyC\Big(R+\frac{y}{2},R-\frac{y}{2}\Big)e^{-ipy}=\int d^Dxd^DyA\Big(R+\frac{y}{2},x)B(x,R-\frac{y}{2}\Big)e^{-ipy}\\
\nonumber =&\int d^Dxd^Dy\frac{d^Dq}{(2\pi )^D}\frac{d^Dk}{(2\pi )^D}A_W\Big(\frac{R}{2}+\frac{y}{4}+\frac{x}{2},q\Big)B_W\Big(\frac{R}{2}-\frac{y}{4}+\frac{x}{2},k\Big)e^{iq(R+\frac{y}{2}-x)}e^{ik(x-R+\frac{y}{2})}e^{-ipy}\\
\nonumber =&\int d^Dz_1d^Dz_2\frac{d^Dq}{(2\pi )^D}\frac{d^Dk}{(2\pi )^D}A_W\Big(R+\frac{z_1}{2},q\Big)B_W\Big(R+\frac{z_2}{2},k\Big)e^{-iqz_2}e^{ikz_1}e^{-ip(z_1-z_2)}\\
\nonumber =&\int d^Dz_1d^Dz_2\frac{d^Dq}{(2\pi )^D}\frac{d^Dk}{(2\pi )^D}A_W(R,q)e^{\frac{z_1}{2}\overset{\leftarrow}{\partial}_R}e^{\frac{z_2}{2}\overset{\rightarrow}{\partial}_R}B_W(R,k)e^{-iqz_2}e^{ikz_1}e^{-ip(z_1-z_2)}\\
\nonumber =&\int d^Dz_1d^Dz_2\frac{d^Dq}{(2\pi )^D}\frac{d^Dk}{(2\pi )^D}A_W(R,q)e^{ikz_1}e^{-\frac{i}{2}\overset{\leftarrow}{\partial}_k\overset{\leftarrow}{\partial}_R}e^{\frac{i}{2}\overset{\rightarrow}{\partial}_q\overset{\rightarrow}{\partial}_R}e^{-iqz_2}B_W(R,k)e^{-ip(z_1-z_2)}\\
\nonumber =&\int\frac{d^Dq}{(2\pi )^D}\frac{d^Dk}{(2\pi )^D}\Big[A_W(R,q)e^{\frac{i}{2}\overset{\rightarrow}{\partial}_k\overset{\leftarrow}{\partial}_R}e^{-\frac{i}{2}\overset{\leftarrow}{\partial}_q\overset{\rightarrow}{\partial}_R}B_W(R,k)\Big]\int d^Dz_1d^Dz_2e^{ikz_1}e^{-iqz_2}e^{-ip(z_1-z_2)}\\
=&A_W(R,p)e^{\frac{i}{2}\overset{\rightarrow}{\partial}_p\overset{\leftarrow}{\partial}_R}e^{-\frac{i}{2}\overset{\leftarrow}{\partial}_p\overset{\rightarrow}{\partial}_R}B_W(R,p)=A_W(R,p)\star B_W(R,p).
\end{align}
We employed the substitutions $z_1=x+\frac{y}{2}-R,\,z_2=x-\frac{y}{2}-R$. If the operator $\hat{\mathcal{O}}$ is invertible, the identity
\begin{align}
O_W(x,p)\star O^{-1}_W(x,p)=1
\end{align}
on phase space is also known as the Groenewold equation of the Weyl-symbol of the operator $\hat{\mathcal{O}}$. Finally note that the Moyal star product is associative but in general not commutative.\par
We will make use of the Groenewold equation to relate the Weyl-symbol of the propagator $G_W$ to the Weyl-symbol of the fermion operator $Q_W$ according to
\begin{align}
Q_W(x,p)\star G_W(x,p)=1
\end{align}
on phase space. The Green function, or propagator, for time translation invariant theories or in thermodynamic equilibrium depends in general only on one (imaginary) time or frequency coordinate, while it depends on two spatial coordinates in spatial configuration, momentum or phase space. We define the Green function matrix element $G$ as the inverse of the fermion operator matrix element $Q$ in configuration and momentum space by
\begin{align}
G(\bold{x},\bold{y},\omega )=\langle \bold{x}|(\omega -\hat{H})^{-1}|\bold{y}\rangle ,\,\,\,\,\tilde{G}(\bold{p},\bold{q},\omega )=\langle \bold{p}|(\omega -\hat{H})|\bold{q}\rangle .
\end{align}
We use bold letters in order to indicate coordinates only referring to spatial directions. $\hat{H}$ denotes the Hamilton operator of the physical theory under consideration. The corresponding Weyl-symbol of the propagator will be denoted by $G_W(\bold{x},\bold{p},\omega )$. The Feynman (or time ordered), Matsubara (or imaginary time ordered), retarded and advanced Green functions or propagators are defined as follows
\begin{align}
\nonumber &G^F_W(\bold{x},\bold{p},\omega )=G_W(\bold{x},\bold{p},\omega +i0^+sgn(\omega )),\,\,\,\, G^M_W(\bold{x},\bold{p},\omega_n )=-iG_W(\bold{x},\bold{p},i\omega_n ),\\
&G^R_W(\bold{x},\bold{p},\omega )=G_W(\bold{x},\bold{p},\omega +i0^+),\,\,\,\,G^A_W(\bold{x},\bold{p},\omega )=G_W(\bold{x},\bold{p},\omega -i0^+)
\end{align}
with the superscripts distinguishing between the four types of propagators in the conventional way. The fermionic Matsubara frequencies are defined by \newline 
$\omega_n=(2n+1)\pi T$ with $n\in\mathbb{Z}$. We denote a positive infinitesimal by $0^+$. We will usually make use of the Matsubara propagator and the associated fermion operator and then leave the corresponding superscript implicit. The superscripts will be reinstated whenever necessary in the corresponding context. The Feynman prescription naturally corresponds to the path integral formulation in Minkowski spacetime, while the Matsubara prescription does so for the path integral formulation in Euclidean space. The retarded propagator relates to causal processes, while the advanced propagator relates to anti-causal processes. Within the Minkowski (Euclidean) path integral of field theory we find for the propagator the following expressions
\begin{align}
G^F(x,y)=-i\langle \psi (x)\bar{\psi}(y)\rangle\,\,\,\,(G^M(x,y)=\langle \psi (x)\bar{\psi}(y)\rangle )
\end{align}
in terms of the Grassmann-valued fermionic field variables $\Psi$ and $\overline{\Psi}$.

\section{The Hall viscosity in the non-relativistic field theory model of Lopez and Fradkin}
\label{SectnonRel}

We present the characteristics of the Hall viscosity within the non-relativistic field theories of electrons and composite fermions for integer and Jain fractional quantum Hall phases. We start out with its general definition as a viscous stress response and represent the stress tensor in terms of fermion operators and associated Weyl symbols. The Hall viscosity is defined within linear response theory and represented by a novel Green function expression assuming Galilean invariance. We then investigate the topological robustness of this expression and calculate it explicitly for non-interacting systems. As a main result we find the non-renormalization of the Hall viscosity by perturbative Coulomb interactions for Galilean invariant fluids. Our discussion draws parallels with previous analysis for the Hall conductivity conducted as well within the composite fermion field theory language and Wigner-Weyl calculus.

\subsection{Preliminaries on the Hall viscosity}
\label{SectHallViscosity}

We now intend to present the definition and a few salient properties of the Hall viscosity for fluids closely following the pedagogical treatment presented in \cite{hoyos2014hall}. The Hall viscosity is a non-dissipative transport coefficient associated with strain rate induced stresses. Small deformations of a (crystalline) medium may be parametrized by a displacement vector $\xi_i$ with $i=1,...,d$ (for $d$ spatial dimensions) which induces a strain $\xi_{ij}=\frac{1}{2}(\partial_i\xi_j+\partial_j\xi_i)$ and a strain rate $\partial_0\xi_{ij}$. The strain rate may be related to a fluid's three velocity $u_i=\partial_0\xi_i\,\Rightarrow\,\partial_iu_j+\partial_ju_i=2\partial_0\xi_{ij}$. Notice that these ``active'' strains due to external forces may also be captured by ``passive'' strains parametrized by metric fluctuations $g_{ij}\to g_{ij}+\delta g_{ij}$ with $\delta g_{ij}=\partial_i\xi_j+\partial_j\xi_i=2\cdot \xi_{ij}$ (notice this factor of two in combination with Eq. (\ref{stressproduction})) corresponding to coordinate transformations $x^i\to x^i+\xi^i$. $\xi$ in this passive description is meant to represent a vector field generating a family of spatial diffeomorphisms. We will be interested only in the case of spatially homogeneous strains (and therefore also homogeneous strain rates). These imply a stress which we parametrize more generally by
\begin{align}
T_{ij}=p\delta_{ij}-\lambda_{ijkl}\xi_{kl}-\eta_{ijkl}\partial_0\xi_{kl}+O((\partial_0)^2).
\label{stressproduction}
\end{align}
The coefficients are the pressure $p$, the elastic modulus $\lambda_{ijkl}=\kappa^{-1}\delta_{ij}\delta_{kl}$ with inverse compressibility $\kappa^{-1}=-V\partial_Vp$ and viscosity tensor $\eta_{ijkl}$. The Hall viscosity forms part of the latter tensor as a scalar coefficient which arises as follows. In a time reversal invariant theory the Onsager relations imply that $\eta_{ijkl}=\eta^S_{ijkl}=\eta^S_{klij}$ which, in the case of rotational invariance, allows for only two distinct transport coefficients, namely the shear viscosity $\eta$ and the bulk viscosity $\zeta$, according to
\begin{align}
\eta_{ijkl}=\eta (\delta_{ik}\delta_{jl}+\delta_{il}\delta_{jk})+\Big(\zeta -\frac{2}{d}\eta \Big)\delta_{ij}\delta_{kl}.
\end{align}
When time reversal invariance is broken, as occurs, e. g., for the Hall fluid characterized by the presence of a strong external magnetic field, the Onsager relations cease to hold in the just presented form and enable the appearance of an antisymmetric viscosity contribution
\begin{align}
\eta_{ijkl}=\eta^S_{ijkl}+\eta^A_{ijkl},\,\,\,\, \eta^A_{ijkl}=-\eta^A_{klij}.
\end{align}
such that $\eta^A_{ijkl}(B)=-\eta^A_{ijkl}(-B)$ if the magnetic field is the only source of time-reversal invariance breaking. It may be proven that the antisymmetric viscosity is dissipationless and does not necessarily vanish in the zero temperature limit (as opposed to the dissipative shear and bulk viscosities). In the case of rotational invariance the antisymmetric viscosity vanishes for $d>2$. This case thus singles out $d=2$ for which an antisymmetric viscosity is allowed if parity is also broken. The so-called Hall viscosity coefficient may then finally be defined by
\begin{align}
\eta^A_{ijkl}=-\frac{\eta_H}{2}(\epsilon_{ik}\delta_{jl}+\epsilon_{jk}\delta_{il}+\epsilon_{il}\delta_{jk}+\epsilon_{jl}\delta_{ik})\,\,\,\,\Rightarrow\,\, \eta_H=-\frac{1}{4}\epsilon_{ik}\delta_{jl}\eta^A_{ijkl}.
\label{hallviscosityextraction}
\end{align}
To sum up the Hall viscosity is a dissipationless transport coefficient parametrizing the strength of stresses induced by nonzero strain rates which in two spatial dimensions is non-vanishing only in theories with time reversal and parity breaking. An example of a fluid allowing for a non-vanishing Hall viscosity is the quantum Hall fluid. Notice that the so defined Hall viscosity should be referred to as geometric Hall viscosity, since we refer here and further below to a definition relative to the metric (see Eqs. (\ref{ZLF0}) and (\ref{ZLF}) below). In general metric and strain may be related in a non-trivial way which can not be captured as presented here \cite{leyva2015generalizing,volovik2015emergent,zubkov2015emergent}. It is then necessary to distinguish them and quantities derived relative to them carefully. This is relevant to, e. g., emergent gravity in graphene \cite{selch2026graphene}.\par
We aim to derive the finite value of the Hall viscosity for the quantum Hall fluid in the mean field approximation of field theory and proof its perturbative non-renormalization by Coulomb interactions in the subsequent sections. \par
Notice that we will from now on assume that the metric has unit determinant $\det (g_{ij})=1$ and is only slightly perturbed from the constant spatial Minkowski metric ($\eta_{ij}=\delta_{ij}$)
\begin{align}
g_{ij}=\delta_{ij}+\delta g_{ij},\,\,\,\,\delta^{ij}\delta g_{ij}=0.
\end{align}
This will be enough to extract the Hall viscosity. Notice that spatial indices may be raised and lowered trivially within the linear response regime.

\subsection{Stress for electrons with Coulomb interactions}
\label{SectStress}

We start from the expression for the Minkowski space partition function of non-relativistic electrons of mass $m$ at chemical potential $\mu$ interacting via Coulomb forces
\begin{equation}
\mathcal{Z} = \int D\psi D\bar{\psi} D \lambda e^{iS[\lambda] + i\int d^3 x \bar{\psi} Q[\lambda] \psi}
\label{Z0}
\end{equation}
with 
\begin{align}
&S[\lambda ]=\frac{1}{2} \int d^3x d^3x^{\prime} \lambda(x) V^{-1}(x,x^{\prime })\lambda(x^\prime),\\
&Q[\lambda]=i\partial_0-A_0(x)+\mu -\lambda(x)+\frac{1}{2m}g^{ij}D_iD_j,\,\,\,\,\bold{D}=\boldsymbol{\partial}+i\bold{A}(x). \label{ZLF0}
\end{align}
The field $\lambda$ is the dynamical density fluctuation (the Hubbard-Stratonovich field) coupling to fermions and mediating Coulomb interactions. $V$ is the (instantaneous/non-retarded) Coulomb potential with inverse $V^{-1}$ and associated configuration space matrix elements $V^{-1}(x,x') = \langle x|V^{-1}|x'\rangle$. The electric charge $e$ contained in $V$ will mostly be set to unity $e=1$. We furthermore employ natural units.\par
The expectation value of the local stress $T_{ij}(x)$ in rotationally invariant systems is given by the variation of the partition function $\mathcal{Z}$ with respect to the external (inverse) metric $g^{ij}(x)$
\begin{align}
\nonumber T_{ij}(x)=&2i\frac{\delta \log \mathcal{Z}}{\delta g^{ij}(x)}=2i\frac{1}{\mathcal{Z}}\frac{\delta \mathcal{Z}}{\delta g^{ij}(x)}=-\frac{2}{\mathcal{Z}}\int D\psi D\bar{\psi} D\lambda e^{iS[\lambda] + i\int d^3 x \bar{\psi}Q[\lambda] \psi}\Big[\frac{\delta}{\delta g^{ij}}S[\lambda]+ \bar{\psi}(x) \frac{\delta}{\delta g^{ij}}Q[\lambda] \psi(x)\Big]. 
\label{fieldtheorystress}
\end{align}
This definition of local stress in the non-relativistic case is motivated in Appendix \ref{appendix0}. At zeroth order of perturbation theory, where $\lambda$ is set to zero, we find 
\begin{align}
T_{ij}(x) =&2i\Tr\Big(\hat{G}[\lambda =0]\frac{\delta\hat{Q}[\lambda =0]}{\delta g^{ij}(x)}\Big)=2i\int\frac{d^3 p}{(2\pi)^3} \Tr\Big[ {G}^{(0)}_W(x,p)\frac{\partial Q^{(0)}_W(x,p)}{\partial g^{ij}}\Big]
\end{align}
where $\hat{Q}^{(0)} \equiv \hat{Q}[0]$ is the fermion operator with interactions disregarded, $\hat{G}^{(0)} = [\hat{Q}^{(0)}]^{-1}$ is the non-interacting Green function.\par
We will denote the interacting Feynman (or later Matsubara) Green function by a calligraphic letter $\hat{\mathcal{G}}$, while by an ordinary letter $\hat{Q}$ we denote the non-interacting fermion operator (without sub- or superscript from now on). Correspondingly, $\hat{\mathcal{Q}}$ is inverse to $\hat{\mathcal{G}}$ while $\hat{G}$ is inverse to $\hat{Q}$. One can see that, including the effects of Coulomb interactions, the stress is to be calculated as the non-renormalized strain operator Weyl-symbol $\frac{\partial}{\partial g^{ij}}Q_W(x,p)$ averaged with the aid of the fully interacting Green function operator Weyl-symbol $\mathcal{G}_W(x,p)$
\begin{equation}
T_{ij}(x)=2i \int \frac{d^3 p}{(2\pi)^3}\Tr \Big[\mathcal{G}_W(x,p)\frac{\partial}{\partial g^{ij}} Q_W(x,p)\Big].
\end{equation}
Notice that this energy momentum tensor computed under the assumption of unit metric determinant implies a restriction to the traceless part of the full energy momentum tensor. We will prove using the results of both \cite{selch2025hall} and \cite{ZZ2021}, relying on Galilean invariance, that, when Coulomb interactions are taken into account perturbatively, the non-renormalized strain operator may be replaced by the renormalized one $\frac{\partial}{\partial g^{ij}}\mathcal{Q}_W(x,p)$. Consequently, we will find the Hall viscosity to be robust under variations of the fine structure constant.  

\subsection{The model by Lopez and Fradkin and its mean field approximation}
\label{SectLFM}

The standard model of electrons interacting via Coulomb forces presented in the previous subsection is equivalent to the model 
defined by the partition function 
\begin{equation}
\mathcal{Z}=\int D\psi D\overline{\psi} D \lambda D\mathcal{A}_{\mu} e^{iS_{CS}[\mathcal{A}]+iS[\lambda] + i\int d^3 x \overline{\psi} Q[\mathcal{A},\lambda] \psi}
\label{Z1}
\end{equation}
with  
\begin{align}
&S[\lambda ]=\frac{1}{2} \int d^3x d^3x^{\prime} \lambda(x)V^{-1}(x,x^{\prime })\lambda(x^\prime),\\
&S_{CS}[\mathcal{A}]=\frac{ \theta}{4} \int d^3x \epsilon^{\mu\nu\rho} \mathcal{A}_{\mu}\mathcal{F}_{\nu\rho},\\
&Q[\mathcal{A},\lambda]=i\partial_0-A_0(x)-\mathcal{A}_0(x)-s_{top}\omega_0(x)+\mu -\lambda(x)+\frac{1}{2m}g^{ij}D_iD_j,\,\,\,\,\bold{D}=\boldsymbol{\partial}+i\bold{A}(x)+i\bold{\mathcal{A}}(x)+is_{top}\boldsymbol{\omega} (x).
\label{ZLF}
\end{align}
This has been argued for in flat spacetime in many particle quantum mechanics in the original work by Lopez and Fradkin (see Appendix A of \cite{Fradkin1991chern}) and from the field theory point of view in Appendices A-C of \cite{selch2025hall} as long as $\theta = 1/(2\pi \,2s)$ and $s$ is an integer. The proof in curved spacetime is presented in Appendix \ref{appendixA}.\par
$S_{CS}$ is a Chern-Simons action comprising the so-called statistical gauge field $\mathcal{A}_{\mu}(x)$ which moreover appears in the covariant derivative of the electrons. The Chern-Simons term is topological. It contributes neither to the energy nor the momentum carried by the fields. The statistical gauge field is emergent and non-perturbative. Lopez and Fradkin integrated out the fermions and the statistical gauge field in their model in an RPA approximation. The residual effective action contains a Chern-Simons term comprising the external electromagnetic gauge field sourced exactly by $S_{CS}$ as given above in terms of the statistical gauge field. The Chern-Simons effective action is the general effective field theory manifestation of the QHE. The strength of the topological term is determined by $\theta = 1/(2\pi \,2s)$, and $s$ is an integer. The Chern-Simons term induces the attachment of $2s$ quanta of magnetic flux (with magnetic flux quantum $\Phi_0=2\pi$ in our units) to each electron. This gives rise to a new effective quasiparticle, the so-called composite fermion. It is subjected to an effective electromagnetic field, the sum of the external field plus the screening statistical field. The former is not confined to the sample, while the latter is. The advantage of the composite fermion formulation is the existence of a mean field theory around which we may expand meaningfully. \par
In a curved spacetime background, the composite fermion acquires a topological spin $s_{top}=s$ due to magnetic flux attachment which couples to an Abelian spin connection $\omega_{\mu}$ \cite{hoyos2012hall} associated with rotations in the 2D plane with components
\begin{align}
\omega_0=\frac{1}{8}\epsilon_{jk}g_{ij}\partial_0g_{ik},\,\,\,\,\omega_i=O(\partial_ig_{jk}).
\end{align}
Note that the temporal component $\omega_0$ is quadratic in deviations of the metric from the flat metric, while the spatial components $\omega_i$ are linear but are nonzero only in the case of spatial inhomogeneity which is why they will be irrelevant for our considerations. To be more precise the topological spin of $s_{top}=s$ may be argued for heuristically invoking composite fermion exchange statistics. The spin polarized and thereby scalar electron does not couple minimally to the spin connection. A composite fermion has a statistical angle of $\pi$ plus $\pi \times 2s$ for $2s$ attached magnetic flux quanta. This is the statistical transmutation evoked by the statistical gauge field via the Chern-Simons term. Therefore it carries a topological spin of s which should couple to the spin connection with its fermionic spin contribution due to spin polarization discarded as for electrons. This can be obtained by composite fermion exchange and the induced Aharonov-Bohm phase.\par
The flux attachment procedure maps a FQHE for electrons to an IQHE for composite fermions. The integer $s$ fixes the statistics of the composite fermion quasiparticles. We intend to continue working with fermions. Other choices of $s$ would lead to Abelian anyons as quasiparticles. It appears that the partition function of Eq. (\ref{Z1}) equals the partition function given by the same expression but without the Abelian statistical gauge field and the spin connection. The chosen values for $\theta$ imply the absence of topological contributions due to the so-called linking number for intertwined particle paths. The advantage of the model by Lopez and Fradkin over that of the previous subsection is that it allows for a mean field approximation where the external electromagnetic fields are screened by the statistical gauge field. The equivalence of the mean field theory model of Lopez and Fradkin to the standard model of electrons, with Coulomb interactions present in both theories, means that a Jain fractional quantum Hall phase for electrons is equivalent to an integer quantum Hall phase for composite fermions with additional topological spin. As a consequence both give rise to the same Hall conductivity and, due to the topological spin for composite fermions, to the same Hall viscosity.\par
We start applying the field theory of Lopez and Fradkin by showing that the mean field theory allows for homogeneous solutions. We will subsequently denote the average particle density by $\bar{\rho}$ and the average current by $\bar{\bold{j}}$. The mean field theory approximation in flat space results in the equations ($g_{ij}=\delta_{ij}\Rightarrow \omega_{\mu}=0$)
\begin{align}
&\frac{\theta}{2}\epsilon^{i\mu\nu} {\cal F}_{\mu\nu}=J^i,\,\,\,\, J^0(z) = \int d^3 z' V^{-1}(z,z')\lambda (z')+\bar{\rho}(z),\,\,\,\, J^{\mu}=(\rho ,\bold{j})=\Big\langle\Big(\overline{\psi} \psi, \overline{\psi} \frac{(-i)\bold{D}}{m} \psi \Big)\Big\rangle .
\label{meanfield0}
\end{align}
A homogeneous flat space solution of these equations reads
\begin{align}
{\theta} {\cal B} = -\bar{\rho},\,\,\,\,{\theta} {\cal E}^i=\epsilon^{ij}\bar{j}^j,\,\,\,\,\lambda(z) = 0,\,\,\,\,\bar{\rho} = \langle\bar{\psi} \psi \rangle ,\,\,\,\,\bar{j}^k = \Big\langle\bar{\psi}\frac{(-i)D^k}{m} \psi \Big\rangle .
\end{align}
The external magnetic field $B$ is screened by the statistical gauge field giving rise to the effective magnetic field $B_{eff}$. If precisely $p$ Landau levels are occupied for $B_{eff}$, then we have $\bar{\rho} = \frac{p}{2\pi}B_{eff}$. We arrive at the equation for the determination of $B_{eff}$
\begin{align}
B_{eff} =&B+\mathcal{B}=B-\frac{\bar{\rho}}{\theta} = B - 2sp B_{eff}\,\,\Leftrightarrow \,\,B_{eff} = \frac{B}{1+2sp}.
\end{align}
A similar situation takes place for the effective electric field
\begin{align}
{E}_{eff}=E+\mathcal{E}=E-\bar{j}/\theta = {E} - 2\pi \, 2s\, \bar{j}
\end{align}
where $\bar{j}$ is the QHE current given by 
\begin{equation}
\bar{j}=\frac{1}{2\pi}\mathcal{N}_{\sigma_H}E_{eff} = \frac{1}{2\pi}\mathcal{N}_{\sigma_H}({E} - 2\pi \, 2s \bar{j})\,.
\label{IQHEforcompfer}
\end{equation}
The average Hall conductivity at zero temperature is then derived to be 
\begin{align}
\sigma_{H} = \frac{1}{2\pi}\frac{\mathcal{N}_{\sigma_H}}{1+2s\mathcal{N}_{\sigma_H}}
\end{align} 
whereby $\bar{j}=\sigma_HE$. The quantity $\mathcal{N}_{\sigma_H}$ is the following expression (to be calculated after Wick-rotation using the Euclidean (Matsubara) Green function at low temperatures) \cite{selch2025hall}
\begin{align}
\nonumber \mathcal{N}_{\sigma_H}=&{\frac{T}{24\pi^2A}} \int d^3x d^3p\epsilon_{\mu\nu\rho}\cdot\Tr\Big[ (G_{eff})_W(x,p)\star \frac{\partial (Q_{eff})_W(x,p)}{\partial p_{\mu}}\star (G_{eff})_W(x,p)\\
&\star \frac{\partial (Q_{eff})_W(x,p)}{\partial p_{\nu}}\star (G_{eff})_W(x,p)\star \frac{\partial (Q_{eff})_W(x,p)}{\partial p_{\rho}}\Big]\,.  
\label{calM4}
\end{align}
$A$ is the area of the sample and $T$ the temperature. The latter is assumed to be small enough in order to allow for a continuum approximation of the discrete Matsubara frequencies $T\ll B_{eff}/m$. In Landau gauge $(Q_{eff})_W (x,p)$ is related to the effective theory Hamiltonian by  
\begin{align}
(Q_{eff})_W(x,p) =i\omega - H_{eff}(p_x,p_y - B_{eff} x).
\end{align}
$(G_{eff})_W$ is the Green function calculated by inversion of $(Q_{eff})_W$. In the mean field approximation the effect of
interactions is included through the effective magnetic field $B_{eff} = B -\rho_0 / \theta$ \cite{Fradkin1991chern} (as well as the effective electric field). Direct calculation gives $\mathcal{N}_{\sigma_H}=p$ (for a derivation see Appendix D of \cite{selch2025hall}). We arrive at the FQHE with Hall conductivity
\begin{equation}
\sigma_{H} = \frac{1}{2\pi}\frac{p}{1+2sp}.
\end{equation}
The screening of the external electric field by the statistical electric field is crucial for the appearance of the FQHE for electrons, while composite fermions experience an IQHE which may be seen from the first equality in Eq. (\ref{IQHEforcompfer}). This reasoning can also be found in \cite{simon1998chern} (see section 3.1 therein). The screening statistical magnetic field appears in the reference frame moving together with the electron liquid. In this fluid reference frame it represents the magnetic fluxes attached to the electrons. Transformation to the laboratory reference frame (where the external electric field is nonzero, and the macroscopic velocity of the electron liquid is nonzero as well) results in a non-trivial transformation of the statistical gauge field fluxes. This is how the screening statistical electric field appears. In particular, the result of this screening is the appearance of an effective fractional value of the electric charge for the electronic quasiparticles $e^{\prime}=e/(1+2sp)$. Composite fermions experience screened electromagnetic fields and respond more weakly than electrons to external fields. Effectively they behave as quasiparticles with charge $e^{\prime}$ relative to external electromagnetic fields.\par
The Hall viscosity is a geometric response function. It has been found that the ratio of Hall viscosity over particle number density is quantized for a translationally and rotationally invariant (Hall) fluid. This is expressed by the relation
\begin{align}
\eta_H=\frac{1}{2}\bar{s}\bar{\rho}=\frac{1}{4}\mathcal{S}\bar{\rho} .
\end{align}
The particle number $N=\bar{\rho}A$ may generally be represented by
\begin{align}
N=\nu N_{\Phi}+\nu\mathcal{S}\frac{\chi}{2}
\end{align}
with filling factor $\nu$, number of flux quanta $AB/\Phi_0$, sample area $A$ and sample shape represented by the Euler characteristic $\chi$ of the spatial Riemann surface. The particle number relation defines the so-called Wen-Zee shift $\mathcal{S}=2\bar{s}$. The quantum number $\bar{s}$ is the so-called average orbital spin. If the external background magnetic field $B$ is large compared to Coulomb interactions, the motion of interacting electrons decouples into Landau orbit motion (which is present in the non-interacting case as well) and guiding center motion (which emerges due to strong electron correlations) \cite{haldane2009hall,haldane2011geometrical,park2014guiding}. This is valid even in the absence of rotational symmetry. The average orbital spin is then composed of the so-called Landau orbit and the guiding center spins. We remark here that we assume Galilean symmetry which is a rather strong assumption. It then directly implies the topological quantization of the Hall viscosity to particle density ratio in our discussion. We will derive a novel Green function expression for the Hall viscosity in the next section which should then inherit this topological character.\par
Notice that we effectively always consider the IQHE either for electrons or composite fermions. The FQHE for electrons is equivalent to the IQHE for composite fermions. We evaluate the Green function expression in Eq. (\ref{calM4}) only for non-interacting systems where it coincides with the first Chern number and thereby, modulo the inverse von Klitzing constant $1/2\pi$, with the quantum Hall response. Notably it also coincides with the Wen-Zee shift for integer quantum Hall phases. This invariant is always an integer valued topological invariant for Green functions regardless of the presence or absence of interactions of the underlying system. It has been studied directly for a number of fractional quantum Hall states in \cite{gurarie2013topological,gurarie2014hall}. The Green function topological expression for fractional quantum Hall phases, which are strongly correlated, follows the Wen-Zee shift rather than the quantized Hall response. Via Wigner-Weyl calculus and the bulk-boundary correspondence this invariant has been rewritten and shown to yield twice the so-called total orbital spin. This identical to the Wen-Zee shift for all fractional quantum Hall states whose wavefunctions may be represented by conformal blocks of associated conformal field theories (including Laughlin, Moore-Read (Pfaffian) and Read-Rezayi (parafermion) states) but might fail for higher states in the Jain hierarchy (of which Laughlin states are the lowest with $p=1$). These findings establish that, for certain fractional quantum Hall states, the Green function expression in Eq. (\ref{calM4}) captures the topology of the Hall viscosity.\par
We will subsequently be concerned with deriving a topological expression for the Hall viscosity of the Hall fluid analogous to $\mathcal{N}_{\sigma_H}$ for the Hall conductivity but different from it. It therefore represents a novel Green function expression which should finally be found to be topological as well for Galilean invariant fluids as motivated shortly before. An external electric field will not be needed which is therefore set to zero and implies a vanishing electric current. There are no persistent currents due to the assumption of rotational invariance. Instead we will assume the presence of a non-vanishing homogeneous stress rate $\partial_0g_{ij}$. We start with the effective mean field theory (whose treatment matches that of a free electron theory in an external electromagnetic field) and derive a topological invariant $\mathcal{N}_{\eta_H}$ in the case of Galilean invariance in the next subsection. Note that the assumption of homogeneity of the spatial metric implies $\omega_i=0$. Subsequently we will explicitly calculate this expression following and slightly adapting the steps outlined in Appendix D of \cite{selch2025hall}. 
\vspace{-0.04cm}
\subsection{Hall viscosity expression in terms of Weyl-symbols of the effective mean field theory operators}
\label{SectTopinvderivation}

In this subsection we provide the derivation of a topological expression for the Hall viscosity in the present context of the mean field approximation featuring spatial homogeneity (which implies $\omega_i=0$) and isotropy (implying vanishing currents and thereby the validity of our expression for the stress tensor) and Galilean boost symmetry (implying the validity of our manipulations of the Green function representation for the Hall viscosity). In the theory by Lopez and Fradkin the Green function in mean field approximation is equal to the free fermion Green function in the presence of the effective magnetic field and in curved space supplemented by a topological spin and a metric  induced spin connection component $\omega_0$. Let us temporarily drop the topological spin induced contributions and assume $s_{top}=0$. The expectation value of the stress in this approximation (in Euclidean space with Euclidean partition function $Z$) is given by 
\begin{align}
\nonumber T_{ij}(x)=&-2\frac{\delta \log Z}{\delta g^{ij}(x)}=-2\frac{1}{Z}\frac{\delta Z}{\delta g^{ij}(x)}=-\frac{2}{Z}\int D\psi D\bar{\psi} D\lambda e^{S[\lambda =0] + \int d^3 x \bar{\psi} Q[\lambda] \psi}\bar{\psi}(x) \frac{\partial}{\partial g^{ij}}Q[\lambda] \psi(x)\\
=&2\Tr\Big[\hat{G}_{eff}\frac{\delta}{\delta g^{ij}(x)}\hat{Q}_{eff}\Big]=2\int \frac{d^3 p}{(2\pi)^3}\Tr\Big[{({G}_{eff})}_W(x,p)\frac{\partial}{\partial g^{ij}} (Q_{eff})_W(x,p)\Big].
\end{align}
Notice that $\hat{Q}_{eff}$ and $\hat{G}_{eff}$ and their associated Weyl symbols contain a nonzero but homogeneous stress rate $\partial_0g_{ij}$. The linear response of the stress to the strain rate is reduced to the response of $\hat{G}_{eff}$ to it. $\hat{G}_{eff}$ ($(G_{eff})_W$) is defined as a solution of the operator (Groenewold) equation
\vspace{-0.02cm}
\begin{align}
\hat{G}_{eff}\hat{Q}_{eff}=&\hat{Q}_{eff}\hat{G}_{eff}=\hat{1}\,\,\,\,((G_{eff})_W \star (Q_{eff})_W=(Q_{eff})_W\star (G_{eff})_W=1).
\end{align}
The expectation value of the (traceless part of the) stress tensor to a homogeneous stress rate within linear response theory averaged over the sample area $A$ at inverse temperature $\beta$ may be calculated to be 
\begin{align}
\nonumber \bar{T}_{ij}=&-\frac{2}{\beta A}\int d^3x\frac{\delta\log Z}{\delta g^{ij}(x)}=\frac{2}{\beta A}\int d^3x\sum_{\omega_n}\Tr\Big[\frac{\delta \hat{Q}^{\omega_n}_{eff}}{\delta g^{ij}(x)}\hat{G}^{\omega_n}_{eff}\Big]\\
\nonumber =&-\frac{2}{\beta A}\int d^3x\sum_{\omega_n}\Tr\Big[\frac{\delta \hat{Q}^{\omega_n}_{eff}}{\delta g^{ij}(x)}\hat{G}^{\omega_n}_{eff}\frac{\partial \hat{Q}^{\omega_n}_{eff}}{\partial (\partial_0g_{kl})}\hat{G}^{\omega_n}_{eff}\Big]\partial_0g_{kl}\\
\nonumber \overset{T\to 0}{=}&-\frac{2}{\beta A}\int d^3x\int \frac{d\omega}{2\pi}\Tr\Big[\frac{\delta \hat{Q}^{\omega}_{eff}}{\delta  g^{ij}(x)}\hat{G}^{\omega}_{eff}\frac{\partial \hat{Q}^{\omega}_{eff}}{\partial (\partial_0g_{kl})}\hat{G}^{\omega}_{eff}\Big]\partial_0g_{kl}\\
\nonumber =&\frac{2}{\beta A}\int d^3x\int \frac{d\omega}{2\pi}\Tr\Big[\frac{\delta \hat{Q}^{\omega}_{eff}}{\delta  g^{ij}(x)}\hat{G}^{\omega}_{eff}\frac{\partial \hat{Q}^{\omega}_{eff}}{\partial g_{kl}}\frac{\partial}{\partial \omega}\hat{G}^{\omega}_{eff}\Big]\partial_0g_{kl}\\
\nonumber =&-\frac{2}{\beta A}\int d^3x\int \frac{d\omega}{2\pi}\Tr\Big[\frac{\delta \hat{Q}^{\omega}_{eff}}{\delta  g^{ij}(x)}\hat{G}^{\omega}_{eff}\frac{\partial \hat{Q}^{\omega}_{eff}}{\partial g_{kl}}\hat{G}^{\omega}_{eff}\frac{\partial \hat{Q}^{\omega}_{eff}}{\partial \omega}\hat{G}^{\omega}_{eff}\Big]\partial_0g_{kl}\\
\nonumber =&-\frac{2}{\beta A}\int d^3x\int \frac{d^3p}{(2\pi)^3}\Tr\Big[\frac{\partial (Q_{eff})_W(x,p)}{\partial  g^{ij}}\star (G_{eff})_W(x,p)\\
&\star\frac{\partial (Q_{eff})_W(x,p)}{\partial g^{kl}}\star (G_{eff})_W(x,p)\star\frac{\partial (Q_{eff})_W(x,p)}{\partial \omega}\star (G_{eff})_W(x,p)\Big]\partial_0g_{kl}.
\label{linearresponseT}
\end{align}
From the second line onwards we use $\delta\hat{Q}_{eff}/\delta g^{ij}(x)$, $\partial\hat{Q}_{eff}/\partial (\partial_0 g^{ij})$ and $\hat{G}_{eff}$ and their associated Weyl-symbols with vanishing strain and strain rate. Our calculations are valid in the linear response regime. In order to reach the fourth line we considered $g_{ij}(t)=g_{ij}(t_0)+(t-t_0)\partial_0g_{ij}|_{t=t_0}+...$ for some reference time $t_0$ such that $g_{ij}(t_0)=\delta_{ij}$. The change of variation with respect to the strain rate $\partial_0g_{ij}$ to variation with respect to $g_{ij}$ is then accompanied by the time operator $t$ which exclusively acts on the rightmost Green function and needs to be transformed into imaginary frequency space.\par
The linear response contribution may be extracted more rigorously within the Wigner-Weyl Green function formalism as follows. We may consider an iterative solution of the Groenewold equation with time dependent fermion bilinear operator $(Q_{eff})_W=(Q_{eff})_W(g_{kl}(t))$ and metric tensor field $g_{kl}(t)=\delta_{kl}+\delta g_{kl}(t)$ with further dependencies suppressed. Linear response refers to an expansion in $\delta g_{kl}(t)$. The viscosity tensor is the stress response to the first time derivative of the metric tensor field. We set $(Q_{eff})_W^{0}=(Q_{eff})_W(\delta_{kl})$ and employ a response expansion in powers of $\delta g_{kl}(t)$ for the propagator Weyl-symbol of the form $(G_{eff})_W(t)=\sum_{n=0}^{\infty}(G_{eff})_W^{n}(t)$, again with suppression of further arguments. At lowest order we set $(G_{eff})_W^{0}\star (Q_{eff})_W^{0}=1$. In the time dependent case the star product has a non-trivial temporal piece which we make explicit via $\star\to\star \circ$ with
\begin{align}
(G_{eff})_W(t)\star \circ (Q_{eff})_W(g_{kl}(t)) = 1 , \quad \circ = e^{\frac{i}{2}\Big((\overleftarrow{\partial}_t)(\overrightarrow{\partial}_{\omega})-(\overleftarrow{\partial}_{\omega})(\overrightarrow{\partial}_t)\Big)},\,\,\,\, \star = e^{\frac{i}{2} \left( (\overleftarrow{\partial}_{x^{i}})(\overrightarrow{\partial}_{p_i})-(\overleftarrow{\partial}_{p_i})(\overrightarrow{\partial}_{x^i})\right)}.
\end{align}
As we are interested in the static limit of the viscous Hall linear response, we may write at first order
\begin{align}
1=\Big[(G_{eff})_W^{1}\star (Q_{eff})_W^{0}+(G_{eff})_W^{0}\star \Big(1+\frac{i}{2}\Big((\overleftarrow{\partial}_t)(\overrightarrow{\partial}_{\omega})-(\overleftarrow{\partial}_{\omega})(\overrightarrow{\partial}_t)\Big)\Big)(Q_{eff})_W(g_{kl}(t))\Big]\Big\rvert_{g_{kl}(t)=\delta_{kl}}.
\end{align}
Multiplication from the right by $\star (G_{eff})_W^{0}$ together with the time independence of $(G_{eff})_W^{0}$ and application of the relation $(G_{eff})_W^{0}\star (Q_{eff})_W^{0}=1$ leads to
\begin{align}
\nonumber (G_{eff})_W^{1}=&\frac{i}{2}\frac{\partial (G_{eff})_W^{0}}{\partial \omega}\star \frac{\partial (Q_{eff})_W^{0}}{\partial g_{kl}}\partial_0g_{kl}\star (G_{eff})_W^{0}\\
=&-\frac{i}{2}(G_{eff})_W^{0}\star\frac{\partial (Q_{eff})_W^{0}}{\partial \omega}\star (G_{eff})_W^{0}\star \frac{\partial (Q_{eff})_W^{0}}{\partial g_{kl}}\star (G_{eff})_W^{0}\partial_0g_{kl}.
\end{align}
Exploiting the antisymmetry $\eta^{A}_{ijkl}=-\eta^{A}_{klij}$ of the Hall viscous contribution a zero temperature limit of the rightmost expression in the first line of in Eq. (\ref{linearresponseT}) expressed in terms of Weyl-symbols with $(G_{eff})_W^{1}$ for the linear approximation of the propagator leads directly to the ultimate equality of Eq. (\ref{linearresponseT}).
As we are working with the mean field effective theory (with the density fluctuation field $\lambda =0$) it follows that
\begin{align}
\frac{\delta (Q_{eff})_W}{\delta g^{ij}}=&\frac{1}{2m}(D_W)_{(i}\star (D_W)_{j)},\,\,\,\,\frac{\partial (Q_{eff})_W}{\partial p_i}=-\frac{i}{m}g^{ij}(D_W)_j\\
\Rightarrow \frac{\delta (Q_{eff})_W}{\delta g^{ij}}=&\frac{i}{2}\frac{\partial (Q_{eff})_W}{\partial p_{(i}}\star (D_W)_{j)}\Big\rvert_{g^{ij}=\delta^{ij}}=\frac{i}{2}(D_W)_{(i}\star\frac{\partial (Q_{eff})_W}{\partial p_{j)}}\Big\rvert_{g^{ij}=\delta^{ij}}.
\label{freetheoryrelations}
\end{align}
with $(D_W)_{\mu}=-ip_{\mu}+iA_{\mu}(x)$. Round brackets on indices indicate normalized symmetrization. These expressions give the following result for the stress averaged over the volume of the system in the linear response approximation (which allow us to set $g_{ij}=\delta_{ij}$ in the subsequent expressions)
\begin{align}
\nonumber \bar{T}_{ij}=&\frac{1}{2\beta A}\int d^3x\int \frac{d^3p}{(2\pi)^3}\Tr\Big[\frac{\partial (Q_{eff})_W(x,p)}{\partial p_{(i}}\star (D_W)_{j)}\star (G_{eff})_W(x,p)\star\frac{\partial (Q_{eff})_W(x,p)}{\partial p_k}\\
&\star (D_W)_l\star (G_{eff})_W(x,p)\star\frac{\partial (Q_{eff})_W(x,p)}{\partial \omega}\star (G_{eff})_W(x,p)\Big]\partial_0g_{kl}.
\label{hallviscositynoninteracting}
\end{align}
The Hall viscosity may then be determined with the aid of Eqs. (\ref{stressproduction}) (note again the factor of two mentioned there) and (\ref{hallviscosityextraction}), respectively. We obtain
\begin{align}
\eta_H=2\cdot \frac{1}{4}\epsilon^{ik}\delta^{jl}\frac{\partial \bar{T}_{ij}}{\partial (\partial_0g^{kl})}=\frac{1}{2\pi}\mathcal{N}_{\eta_H}B_{eff}
\end{align}
with
\begin{align}
\nonumber \mathcal{N}_{\eta_H}=&\frac{1}{16\pi^2\beta A}\int d^3x\int d^3p\frac{1}{B_{eff}}\epsilon^{ik}\delta^{jl}\Tr\Big[\frac{\partial (Q_{eff})_W(x,p)}{\partial p_i}\star (D_W)_j\star (G_{eff})_W(x,p)\\
&\star\frac{\partial (Q_{eff})_W(x,p)}{\partial p_k}\star (D_W)_l\star(G_{eff})_W(x,p)\star \frac{\partial (Q_{eff})_W(x,p)}{\partial \omega}\star (G_{eff})_W(x,p)\Big].
\label{generaltopologicalterm}
\end{align}
This is the general linear response expression for the case $s_{top}=0$. Explicit symmetrization of indices before index contraction in Eq. (\ref{generaltopologicalterm}) is not necessary, as we have verified by explicit calculation. We calculate $\mathcal{N}_{\eta_H}$ as presented in Eq. (\ref{generaltopologicalterm}) explicitly in Appendix \ref{appendixC}.\par
Translational invariance modulo a vector potential reduces the phase space dependence $(x,p)$ of the fermion operator and its inverse to $i(D_W)_{\mu}=p_{\mu}-A_{\mu}(x)$. We assume that $A_{\mu}$ only comprises a constant (effective) background magnetic field. Notice that we may replace exactly one Moyal star product in an integrand over phase space by an ordinary product for a boundaryless phase space due to the application of integration by parts. This implies the trace structure as applied to the Weyl symbols. The Moyal star products may not be removed entirely in the presence of a homogeneous magnetic field, since not all canonical momenta are good quantum numbers. This is manifest in the resulting Moyal star product representation
\begin{align}
\star = \exp\Big(\frac{i}{2}B\epsilon_{ij}\overset{\leftarrow}{\partial}_{(D_W)_i}\overset{\rightarrow}{\partial}_{(D_W)_j}\Big).
\end{align}
The additional assumption of rotational invariance implies that $(Q_{eff})_W$ and its inverse are in general functions of $(\omega ,S)$ where $\omega$ is frequency and $S=(D_W)_i\star g^{ij}\star (D_W)_j$. Notice that a theory with generic fermion bilinear operator Weyl symbol
\begin{align}
Q_W(\omega ,S)=(Q_{eff})_W(\omega ,S)+\Delta Q_W(\omega ,S)
\end{align}
will not lead to the Green function expression presented in Eq. (\ref{generaltopologicalterm}), as the relations in Eq. (\ref{freetheoryrelations}) are only valid for Galilean invariant systems. In the absence of Galilean boost symmetry (allowing for $\Delta Q_W(\omega ,S)\neq 0$) Eq. (\ref{generaltopologicalterm}) coincides with the Hall viscosity only up to subleading terms in the filling factor. This is explained in detail in Appendix \ref{appendixD}.
The calculation of $\mathcal{N}_{\eta_H}$ in Appendix \ref{appendixC} proceeds rather generally (but assumes absence of interactions) in such a way that only eigenfunctions (of $S|_{g^{ij}=\delta^{ij}}$) and energy eigenvalues of the (effective) Hamiltonian in a two Landau level neighborhood of the chemical potential are needed as input to evaluate the Hall viscosity. We did not succeed in proving the topological invariance of $\mathcal{N}_{\eta_H}$ in Eq. (\ref{generaltopologicalterm}) from the Green function expression directly in a manner analogous to $\mathcal{N}_{\sigma_H}$ via infinitesimal variation. Topological stability of $\mathcal{N}_{\eta_H}$ under infinitesimal variation $Q_W\to Q_W+\delta Q_W$ for $Q_W$ is only found for the leading contribution in the filling factor $\nu$ (which is of order $p^2$ for $p$ filled (effective) Landau levels) as explained in Appendix \ref{appendixE}. If we perform a variation in the subspace of non-interacting, Galilean invariant theories such that the energy levels in the two Landau level neighborhood of the chemical potential remain equidistant relative to each other and the eigenfunctions are unmodified, we find invariance. This may be seen directly from Eq. (\ref{hallviscosityintermediate}). Finally the invariant of Eq. (\ref{generaltopologicalterm}) is in one-to-one correspondence with Haldane's Landau orbit spin contribution to the Hall viscosity.

In order to be able to extract the full Hall viscosity for Jain fractional quantum Hall states we will need to consider what happens when $s_{top}\neq 0$. Spatial homogeneity implies that $\omega_i=0$, while $\omega_0\neq 0$ is possible. In the linear response regime this modifies our previous calculations by the addition of a term
\begin{align}
\Delta^{s_{top}}T_{ij}(x)=2\Tr\Big[\hat{G}_{eff}\frac{\delta^2}{\delta g^{ij}(x)\delta (\partial_0g_{kl})}\hat{Q}_{eff}\Big]\partial_0g_{kl}
\end{align}
which induces a shift in the Hall viscosity $\eta_H\to\eta_H+\Delta^{s_{top}}\eta_H$ by the explicit amount
\begin{align}
\Delta^{s_{top}}\eta_H=&2\cdot \frac{1}{4}\epsilon^{ik}\delta^{jl}\frac{\partial \Delta^{s_{top}}\bar{T}_{ij}}{\partial (\partial_0g^{kl})}=-\frac{s}{2}\lim_{T\to 0}\sum_{n=0}^{\infty}\sum_{\omega_m}\Tr\Big[(G_{eff})^{\omega_m}_n\Big]=\frac{1}{2\pi}(\Delta^{s_{top}}\mathcal{N}_{\eta_H})B_{eff}
\label{shifthallviscosity}
\end{align}
The sums are over Landau levels $n$ and Matsubara frequencies $\omega_m$, respectively. This shift also represents a topological term so long as the homogeneous particle number density is unaffected by perturbations which we will assume throughout subsequent considerations. We will argue in Appendix \ref{appendixC} that for composite fermions filling $p$ effective Landau levels (with the spatial momentum integral $\int \frac{d^2\bold{p}}{(2\pi )^2}$ quenched to a discrete sum of Landau levels by the background magnetic field)
\begin{align}
\lim_{T\to 0}\sum_{n=0}^{\infty}\sum_{\omega_m}\Tr [(G_{eff})^{\omega_m}_n]=-p\frac{B_{eff}}{2\pi}\Rightarrow\,\,\Delta^{s_{top}}\eta_H=\frac{sp}{2}\frac{B_{eff}}{2\pi},\,\,\Delta^{s_{top}}\mathcal{N}_{\eta_H}=\frac{sp}{2}.
\label{shifttopologicalspin}
\end{align}
The additional contribution to the Hall viscosity stands in one-to-one correspondence with Haldane's guiding center spin. 

\subsection{Hall viscosity expression in terms of Weyl-symbols of the fully interacting field theory operators}
\label{SectTopinvderivation2}

We come to the derivation of the Green function expression for the Hall viscosity in the fully interacting field theory featuring Galilean invariance. We omit for now the topological spin part in the case of the Jain fractional Hall phases. This implies a renormalization of the fermion bilinear operator $\hat{Q}_{eff}\to \hat{\mathcal{Q}}=\hat{Q}_{eff}-\hat\Sigma$ and the Green function $\hat{G}_{eff}\to\hat{\mathcal{G}}=\hat{\mathcal{Q}}^{-1}$. $\hat\Sigma$ is the self-energy operator. The expectation value of the (traceless part of the) stress tensor (in Euclidean space with Euclidean partition function $Z$) is given by 
\begin{align}
T_{ij}(x)=-2\frac{\delta \log Z}{\delta g^{ij}(x)}=-2\frac{1}{Z}\frac{\delta Z}{\delta g^{ij}(x)}=-\frac{2}{Z}\int D\psi D\bar{\psi} D\lambda e^{S[\lambda] + \int d^3 x \bar{\psi} Q[\lambda] \psi}\bar{\psi}(x) \frac{\partial}{\partial g^{ij}}Q[\lambda] \psi(x)+\Delta^{\lambda}T_{ij}(x). 
\label{stresstensorfermionic}
\end{align}
The additional term $\Delta^{\lambda}T_{ij}(x)$ may be induced by a metric dependence of $S[\lambda ]$ which we will come to shortly. For now we restrict attention to the fermionic contribution and suppress $\Delta^{\lambda}T_{ij}$. Within operator and Weyl-symbol notation this implies for the fully interacting case
\begin{align}
T_{ij}(x)=2\Tr\Big[\hat{\mathcal{G}}\frac{\delta}{\delta g^{ij}(x)}\hat{Q}_{eff}\Big]=2\int \frac{d^3 p}{(2\pi)^3}\Tr\Big[{\mathcal{G}}_W(x,p)\frac{\partial}{\partial g^{ij}} (Q_{eff})_W(x,p)\Big].
\end{align}
The mean field operator $\hat{Q}_{eff}$ and fully interacting $\hat{\mathcal{G}}$ contain a nonzero but homogeneous stress rate $\partial_0g_{ij}$. For conciseness we will use the short-hand notation $\hat{Q}_{eff}=\hat{Q}$ and similarly for the associated Weyl-symbol. The response of the stress to the stress rate is reduced to the response of $\hat{\mathcal{G}}$ to it. Notice that 
\begin{align}
\nonumber \frac{\partial \mathcal{Q}_W(x,p)}{\partial g^{kl}}=&\frac{\partial Q_W(x,p)}{\partial g^{kl}}-\frac{\partial \Sigma_W(x,p)}{\partial g^{kl}}\\
\nonumber =&\frac{i}{2}\frac{\partial Q_W(x,p)}{\partial p_{(k}}\star (D_W)_{l)}-\frac{\partial \Sigma_W(x,p)}{\partial g^{kl}}\\
=&\frac{i}{2}\frac{\partial \mathcal{Q}_W(x,p)}{\partial p_{(k}}\star (D_W)_{l)}+\frac{i}{2}\frac{\partial \Sigma_W(x,p)}{\partial p_{(k}}\star (D_W)_{l)}-\frac{\partial \Sigma_W(x,p)}{\partial g^{kl}}
\end{align}
Assume now that the self-energy is a functional of the kinetic energy and the Coulomb potential. Then
\begin{align}
\frac{i}{2}\frac{\partial \Sigma_W(x,p)}{\partial p_{(k}}\star (D_W)_{l)}-\frac{\partial \Sigma_W(x,p)}{\partial g^{kl}}=\Big(\frac{\delta \Sigma}{\delta V}\frac{\partial V}{\partial g^{kl}}\Big)_W(x,p)+...\,\,\,\,\Big(\Sigma =\Sigma \Big(-g^{ij}\frac{D_iD_j}{2m},V(\mathcal{D}[0,x])\Big)\Big).
\end{align}
Importantly additional terms subleading in the filling factor have been omitted. This will be a common practice within this subsection and arises because a relation analogous to Eq. (\ref{freetheoryrelations}) is not generally valid. The Coulomb potential depends on the (inverse) metric via the geodesic distance
\begin{align}
\mathcal{D}[x,y]=\int_0^1d\xi\sqrt{g_{\mu\nu}(\bold{r}(\xi ))\frac{d\bold{r}^{\mu}}{d\xi}\frac{d\bold{r}^{\nu}}{d\xi}}
\end{align}
which is the length of the geodesic $\bold{r}(\xi )$ satisfying $\bold{r}(0)=x$, $\bold{r}(1)=y$. Differentiation of the geodesic distance with respect to a spatially homogeneous metric yields (see, e. g., \cite{bradlyn2012kubo})
\begin{align}
\frac{\partial}{\partial g_{\mu\nu}}\mathcal{D}[0,x]\Big\rvert_{g_{\mu\nu}=\eta_{\mu\nu}}=\frac{1}{2}\frac{x^{\mu}x^{\nu}}{|x|},\,\,\,\,\mathcal{D}[0,x]\Big\rvert_{g_{\mu\nu}=\eta_{\mu\nu}}=|x|\,\,\,\,\Rightarrow\,\,\frac{\partial}{\partial g_{\mu\nu}}V(\mathcal{D}[0,x])=-\frac{1}{2}\frac{1}{4\pi}\frac{x^{\mu}x^{\nu}}{|x|^3}.
\end{align}
We will need the differentiation with respect to the inverse metric which will yield an additional minus sign for the spatial components. Consequently the expectation value of the stress tensor due to a homogeneous strain rate in linear response theory averaged over the sample area $A$ at inverse temperature $\beta$ in the fully interacting case can be calculated to be 
\begin{align}
\nonumber \bar{T}_{ij}=&-\frac{2}{\beta A}\int d^3x\int \frac{d^3p}{(2\pi)^3}\Tr\Big[\frac{\partial Q_W(x,p)}{\partial g^{ij}}\star \mathcal{G}_W(x,p)\star\frac{\partial \mathcal{Q}_W(x,p)}{\partial g^{kl}}\\
\nonumber &\star \mathcal{G}_W(x,p)\star\frac{\partial \mathcal{Q}_W(x,p)}{\partial \omega}\star \mathcal{G}_W(x,p)\Big]\partial_0g_{kl}\\
\nonumber =&\frac{-i}{\beta A}\int d^3x\int \frac{d^3p}{(2\pi)^3}\Tr\Big[\frac{\partial Q_W(x,p)}{\partial g^{ij}}\star \mathcal{G}_W(x,p)\star\frac{\partial \mathcal{Q}_W(x,p)}{\partial p_k}\star (D_W)_l\\
\nonumber &\star \mathcal{G}_W(x,p)\star\frac{\partial \mathcal{Q}_W(x,p)}{\partial \omega}\star \mathcal{G}_W(x,p)\Big]\partial_0g_{kl}\\
\nonumber &-\frac{2}{\beta A}\int d^3x\int \frac{d^3p}{(2\pi)^3}\Tr\Big[\frac{\partial Q_W(x,p)}{\partial  g^{ij}}\star \mathcal{G}_W(x,p)\star\Big[\frac{i}{2}\frac{\partial \Sigma_W(x,p)}{\partial p_k}\star (D_W)_l-\frac{\partial \Sigma_W(x,p)}{\partial g^{kl}}\Big]\\
\nonumber &\star \mathcal{G}_W(x,p)\star\frac{\partial \mathcal{Q}_W(x,p)}{\partial \omega}\star \mathcal{G}_W(x,p)\Big]\partial_0g_{kl}\\
\nonumber =&\frac{1}{2\beta A}\int d^3x\int \frac{d^3p}{(2\pi)^3}\Tr\Big[\frac{\partial Q_W(x,p)}{\partial p_{(i}}\star (D_W)_{j)}\star \mathcal{G}_W(x,p)\star\frac{\partial \mathcal{Q}_W(x,p)}{\partial p_k}\star (D_W)_l\\
\nonumber &\star \mathcal{G}_W(x,p)\star\frac{\partial \mathcal{Q}_W(x,p)}{\partial \omega}\star \mathcal{G}_W(x,p)\Big]\partial_0g_{kl}\\
\nonumber &-\frac{2}{\beta A}\int d^3x\int \frac{d^3p}{(2\pi)^3}\Tr\Big[\frac{(D_W)_{(i}\star (D_W)_{j)}}{2m}\star \mathcal{G}_W(x,p)\star\Big[\frac{\delta \Sigma_W(x,p)}{\delta V}\frac{\partial}{\partial g^{kl}}V\Big]\\
&\star \mathcal{G}_W(x,p)\star\frac{\partial \mathcal{Q}_W(x,p)}{\partial \omega}\star \mathcal{G}_W(x,p)\Big]\partial_0g_{kl}+...
\label{hallviscosityinteracting}
\end{align}
\red{Round brackets around indices are again employed to denote normalized symmetrization.} Consider the first term in the final equality of Eq. (\ref{hallviscosityinteracting}). It stands in one-to-one correspondence with the non-interacting theory expression of Eq. (\ref{hallviscositynoninteracting}) up to interaction corrections of Weyl symbols of the propagator and its inverse excluding the leftmost fermion operator Weyl symbol. An analogous conclusion has been made for the Hall conductivity \cite{selch2023hall}. The physical expression for the Hall viscosity with inclusion of perturbative interactions seems to deviate from the topological expression due to the appearance of the non-interacting theory fermion operator.
We label the second term in the final equality of Eq. (\ref{hallviscosityinteracting}) $\Delta^{\Sigma}\bar{T}_{ij}$. It may be written as
\begin{align}
\nonumber \Delta^{\Sigma}\bar{T}_{ij}=&-\frac{2}{\beta A}\int d^3x\int \frac{d^3p}{(2\pi)^3}\Tr\Big[\frac{(D_W)_{(i}\star (D_W)_{j)}}{2m}\star \mathcal{G}_W(x,p)\star\Big[\int d^3z\frac{\delta \Sigma_W(x,p)}{\delta V(|z|)}\Big(-\frac{1}{2}\frac{1}{4\pi}\frac{z^kz^l}{|z|^3}\Big)\Big]\\
&\star \mathcal{G}_W(x,p)\star\frac{\partial \mathcal{Q}_W(x,p)}{\partial \omega}\star \mathcal{G}_W(x,p)\Big]\partial_0g_{kl}.
\end{align} 
Notice that the averaged (traceless part of the) energy momentum tensor may pick up another piece in the interacting case due to the variation of $S[\lambda ]$ which we denoted by $\Delta^{\lambda}\bar{T}_{ij}$.\par
On the one hand the density fluctuation field $\lambda$ couples the Coulomb interaction to the fermions thereby renormalizing the fermion operator and Green function. On the other hand the bosonic action $S[\lambda ]$ itself may contribute to the (traceless part of the) energy momentum tensor according to
\begin{align}
\nonumber \Delta^{\lambda}\bar{T}_{ij}=&-\frac{2}{Z}\int D\psi D\bar{\psi} D\lambda e^{S[\lambda]+\int d^3 x \bar{\psi}Q[\lambda] \psi}\int d^3zd^3z^{\prime} \lambda (z)\Big[\frac{\partial}{\partial g^{ij}}\frac{1}{2}V^{-1}(z,z^{\prime})\Big]\lambda (z^{\prime})\\
\nonumber =&-\frac{2}{Z}\int D\psi D\bar{\psi} D\lambda e^{S[\lambda]+\int d^3 x \bar{\psi} Q[\lambda] \psi}\int d^3zd^3z^{\prime} \lambda (z) \Big[-\frac{1}{4}\frac{1}{4\pi}\frac{(z-z^{\prime})^{\langle i}(z-z^{\prime})^{j\rangle}}{|z-z^{\prime}|}\Big]\lambda (z^{\prime})\\
=&-\frac{2V}{Z}\int D\psi D\bar{\psi} D\lambda e^{S[\lambda]+\int d^3 x \bar{\psi} Q[\lambda] \psi}\int d^3z\lambda (z) \Big[-\frac{1}{4}\frac{1}{4\pi}\frac{z^{\langle i}z^{j\rangle}}{|z|}\Big]\lambda (0).
\label{bosonicstresstensorpiece}
\end{align}
We used translational invariance to reach the final equality of Eq. (\ref{bosonicstresstensorpiece}). Notice that both $\Delta^{\Sigma}\bar{T}_{ij}$ and $\Delta^{\lambda}\bar{T}_{ij}$ contain an integration of the form
\begin{align}
\int d^3zF(|z|)z^{\langle i}z^{j\rangle}
\end{align}
for some function $F(|z|)$ where $\langle ...\rangle$ projects out the traceless part. The coordinate $z$ is a relative coordinate such that for any sample shape the integration region is symmetric around $z=0$. It may therefore be concluded that the contributions $\Delta^{\Sigma}\bar{T}_{ij}$ and $\Delta^{\lambda}\bar{T}_{ij}$ to the (traceless part of the) stress tensor in Eq. (\ref{bosonicstresstensorpiece}) vanish identically. The Hall viscosity may then be determined as before in the absence of interactions
\begin{align}
\eta_H=2\cdot\frac{1}{4}\epsilon^{ik}\delta^{jl}\frac{\partial \bar{T}_{ij}}{\partial (\partial_0g^{kl})}=\frac{1}{2\pi}\mathcal{N}_{\eta_H}B_{eff}
\end{align}
but now with
\begin{align}
\nonumber \mathcal{N}_{\eta_H}=&\frac{1}{16\pi^2\beta A}\int d^3x\int d^3p\frac{\epsilon^{ik}\delta^{jl}}{B_{eff}}\Tr\Big[\frac{\partial Q_W(x,p)}{\partial p_{(i}}\star (D_W)_{j)}\star \mathcal{G}_W(x,p)\star\frac{\partial \mathcal{Q}_W(x,p)}{\partial p_{(k}}\star (D_W)_{l)}\\
&\star \mathcal{G}_W(x,p)\star\frac{\partial \mathcal{Q}_W(x,p)}{\partial \omega}\star \mathcal{G}_W(x,p)\Big].
\end{align}
The topological invariance of $\mathcal{N}_{\eta_H}$ is not obvious in the interacting case. We intend to reduce the proof of the topological nature of $\mathcal{N}_{\eta_H}$ to the same methods which have been used for $\mathcal{N}_{\sigma_H}$. We especially need to show that
\begin{align}
\nonumber\delta=&\frac{1}{16\pi^2\beta A}\int d^3x\int d^3p\frac{\epsilon^{ik}\delta^{jl}}{B_{eff}}\Tr\Big[\frac{\partial \Sigma_W(x,p)}{\partial p_{(i}}\star (D_W)_{j)}\star \mathcal{G}_W(x,p)\star\frac{\partial \mathcal{Q}_W(x,p)}{\partial p_{(k}}\star (D_W)_{l)}\\
&\star \mathcal{G}_W(x,p)\star\frac{\partial \mathcal{Q}_W(x,p)}{\partial \omega}\star \mathcal{G}_W(x,p)\Big]=0.
\label{deviationfromtopologicalexpression}
\end{align}
\subsection{Proof of Hall viscosity topological invariance of Galilean invariant (Hall) fluids}
\label{Sectnonrenormalization}

Up to now we reached to showing that the condition of Eq. (\ref{deviationfromtopologicalexpression}) is necessary in order to proof the topological invariance of the Hall viscosity in the presence of Coulomb interactions disregarding the topological spin induced correction to the Hall viscosity for fractional quantum Hall fluids. The perturbative non-renormalization theorem proven in \cite{ZZ2021} allows us to draw conclusions on the averaged stress tensor in the interacting theory as follows.\par
By application of a reference frame transformation by a Galilean boost to the Hall fluid frame (H) from the laboratory frame (L) we found that the non-renormalization of the averaged electric current in the latter frame may be proven from that of the averaged density in that of the former. With the identity $J=J^H_{\mu}dx_H^{\mu}=J^{L}_{\mu}dx_L^{\mu}$ ($\mu =0,1,2$) such that $J_{\mu}=(\rho ,\bold{j})$ and $t=t_H=t_L$, $\bold{x}_H=\bold{x}_L-\bold{v}_Bt$ we obtained \cite{selch2025hall}
\begin{align}
\nonumber \bar{\rho}_L=&-\frac{1}{\beta A}\int d^3x\frac{\delta\log Z}{\delta A_0(x)}=\frac{1}{\beta A}\int d^3x \int \frac{d^3p}{(2\pi )^3}\Tr\Big[(-i)\mathcal{G}_W(x,p)\star \frac{\partial Q_W(x,p)}{\partial p_3}\Big]=\bar{\rho}_H\\
\bar{j}_L^k=&-\frac{1}{\beta A}\int d^3x\frac{\delta\log Z}{\delta A_k(x)}=\frac{1}{\beta A}\int d^3x \int \frac{d^3p}{(2\pi )^3}\Tr\Big[\mathcal{G}_W(x,p)\star \frac{\partial Q_W(x,p)}{\partial p_k}\Big]=\bar{\rho}_Lv_B^k,\,\,\,\,\bar{j}_H^k=0.
\end{align}
The boost velocity $\bold{v}_B$ has to be such as to cancel out the Lorentz force $\bold{F}_L(\bold{v}_B)=0$. Therefore $\bar{\bold{j}}_L$ is not perturbatively renormalized by Coulomb interactions if and only if $\bar{\rho}_L=\bar{\rho}_H$ is not.\par 
The Hall viscosity is considered in the natural rest frame of the fluid which we refer to here also as the Hall fluid frame. It is assumed to be rotationally invariant and therefore neither an energy current nor an electric current are present. We consider a boost of this system to a frame (L) with velocity $\bold{v}_B$ such that $\bold{F}_L(-\bold{v}_B)=0$ which will naturally give rise to a \red{momentum density} $(\bar{T}_L)^0_{\,\,\,k}$ accounted for by the boosted energy density and stress according to $(\bar{T}_L)^0_{\,\,\,i}=-m\bar{\rho}_H v_B^i-v_B^k(\bar{T}_H)^k_{\,\,\,i}$. By Galilean invariance this \red{momentum density} is proportional to the electric current by a factor $\frac{m}{e}$ with effective electron mass $m$ and charge $e$. This may be seen by consulting Eq. (\ref{chargesandcurrents}) and imposing flat spacetime conditions. Since we know that the electric current, and therefore also electric charge, is not perturbatively renormalized we can conclude the same for the energy current and then the stress independent of the boost direction due to rotational invariance of the Hall fluid frame. More explicitly we have
\begin{align}
\nonumber (\bar{T}_L)^0_{\,\,\,k}=m\bar{j}_L^k=&\frac{m}{\beta A}\int d^3x \int \frac{d^3p}{(2\pi )^3}\Tr\Big[\mathcal{G}_W(x,p)\star \frac{\partial Q_W(x,p)}{\partial p_k}\Big]\\
=&-\frac{i}{\beta A}\int d^3x \int \frac{d^3p}{(2\pi )^3}D_W^k\Tr\Big[\mathcal{G}_W(x,p)\star \frac{\partial Q_W(x,p)}{\partial p_3}\Big].
\end{align}
From here on we may argue diagrammatically analogous to the case of charge density non-renormalization \cite{selch2025hall}. Alternatively we may consider the stress tensor directly following Eq. (\ref{freetheoryrelations})
\begin{align}
\nonumber (\bar{T}_H)_{ij}=&\frac{2}{\beta A}\int d^3x \int \frac{d^3p}{(2\pi )^3}\Tr\Big[\mathcal{G}_W(x,p)\star \frac{\delta Q_W(x,p)}{\delta g_{ij}}\Big]\\
\nonumber =&\frac{1}{\beta A}\int d^3x \int \frac{d^3p}{(2\pi )^3}i(D_W)_{(j}\Tr\Big[\mathcal{G}_W(x,p)\star \frac{\partial Q_W(x,p)}{\partial p_{i)}}\Big]\\
=&\frac{1}{\beta A}\int d^3x \int \frac{d^3p}{(2\pi )^3}i(D_W)_{(j}\Tr\Big[\mathcal{G}_W(x,p)\star \frac{\partial}{\partial p_{i)}}(\mathcal{Q}_W(x,p)-\Sigma_W(x,p))\Big].
\end{align}
From here on we may argue diagrammatically analogous to the case of current density non-renormalization. This is possible exactly for the traceless part of the flat space stress tensor. The diagrammatic proof relates the perturbatively renormalized stress at each loop order to a sum of so-called progenitor diagrams (Feynman diagrams without external current insertion) supplemented by a total
derivative operator inserted at the left-hand side into the momentum integral of a fermion loop identical to the loop momentum. This equality has been achieved in such a way as to cancel symmetry factors present in the sum of progenitor diagrams. Symmetry factors are absent after stress insertions into the progenitor diagrams turning them into contributions to the stress. The index combinations relevant to extract the Hall viscosity imply the momentum integral referred to to vanish identically. This implies the perturbative non-renormalization of the stress by Coulomb interactions in the Hall fluid frame and thereby $\delta =0$.\par
We now discuss the fractional quantum Hall Jain states. Recall that we found for the shift of the stress tensor and the Hall viscosity induced by the topological orbital spin
\begin{align}
\nonumber \Delta^{s_{top}}T_{ij}(x)=&2\Tr\Big[\hat{G}_{eff}\frac{\delta^2}{\delta g^{ij}(x)\delta (\partial_0g_{kl})}\hat{Q}_{eff}\Big] \partial_0g_{kl},\\
\Delta^{s_{top}}\eta_H=&2\cdot\frac{1}{4}\epsilon^{ik}\delta^{jl}\frac{\partial \Delta^{s_{top}}\bar{T}_{ij}}{\partial (\partial_0g^{kl})}=-\frac{s}{2}\int \frac{d^3p}{(2\pi )^3}\Tr\Big[G_{eff}(p)\Big].
\end{align}
Hence for the fractional quantum Hall fluid we need to consider the renormalization of the momentum space integral of the propagator trace. Infinitesimally we find
\begin{align}
\nonumber\delta \Delta^{s_{top}}\eta_H=&-\frac{s}{2}\int \frac{d^3p}{(2\pi )^3}\Tr\Big[\delta \tilde{G}_{eff}(p)\Big]=-\frac{s}{2}\int \frac{d^3p}{(2\pi )^3}\Tr\Big[\frac{\delta \tilde{G}_{eff}(p)}{\delta \tilde{\Sigma} (p)}\delta\tilde{\Sigma} (p)\Big]\\
=&-\frac{s}{2}\int \frac{d^3p}{(2\pi )^3}\Tr\Big[i\frac{\partial \tilde{G}_{eff}(p)}{\partial p_3}\delta\tilde{\Sigma} (p)\Big]=\frac{s}{2}\int \frac{d^3p}{(2\pi )^3}\Tr\Big[i\tilde{G}_{eff}(p)\frac{\partial\delta\tilde{\Sigma} (p)}{\partial p_3}\Big].
\label{deltasigmapiece}
\end{align}
The final expression in Eq. (\ref{deltasigmapiece}) is proportional to the infinitesimal charge density shift induced by the self-energy \cite{selch2025hall} which vanishes. Finite self-energies will still not change the Hall viscosity shift either, since the charge density is independent of perturbative Coulomb interactions.\par
This finally proves the claim of perturbative non-renormalization of the Hall viscosity by Coulomb interactions valid for both integer and fractional Jain quantum Hall states.

\section{Discussion}
\label{SectDiscussion}

The main objective of this work has been to elucidate the topological robustness and non-renormalization under perturbative Coulomb interactions of the Hall viscosity for integer and Jain fractional quantum Hall states from the point of view of Green functions. In the couse of our work we introduced a novel Green function representation of the Hall viscosity. This is explicitly distinct from earlier work by V. Gurarie and collaborators who consider the well-known Green function invariant, which for non-interacting electrons coincides with the first Chern number and which we labeled $\mathcal{N}_{\sigma_H}$, for several fractional quantum Hall states which comprise strongly correlated electrons. It may therefore analogously be interesting to investigate which values our Green function expression, labeled $\mathcal{N}_{\eta_H}$, will take when evaluated for fractional quantum Hall states. The conditions required to ensure a topological character of the Hall viscosity for the integer and fractional Jain quantum Hall fluids are translational and rotational invariance of these phases of matter which has been established in existing literature. Our novel Green function expression as well as our non-renormalization proof rely on full Galilean invariance. This implies the topology of the Hall viscosity and our Green function expressions, even though we have not been able to prove it via infinitesimal variation. These symmetry conditions are only relaxed insofar as the homogeneous external magnetic field derives from a non-homogeneous vector potential. This spatial dependence needs to be weak (implying $Ba^2\ll 1$) in order to ensure that momentum space matrix elements are close to being diagonal and the approximate Wigner-Weyl continuum theory calculus to be valid precisely (its range of validity coincides with that of effective field theory). A system close to homogeneity generically motivates the use of the continuum field theory Wigner-Weyl calculus within which a number of essential manipulations can be performed straightforwardly as shown explicitly in previous sections. In this way we may simultaneously keep track of non-trivial ordering of operators which descends to the non-commutative algebra of Weyl-symbols. This is crucial to capture the fact that momenta cease to be a good quantum numbers (at least one of them) in the presence of a homogeneous magnetic field.\par
The general expression for the Hall viscosity in the Green function representation as derived from linear response theory has been given in Eqs. (\ref{generaltopologicalterm}) and (\ref{shifthallviscosity}), respectively. We then exploited several properties of the system under consideration. First the (almost) homogeneity of the phase ensures that the Weyl symbol of the covariant derivative is linear in phase space variables. Second the Green function and its inverse are functions of the covariant derivative only. Third due to rotational invariance this dependence is on the squared covariant derivative. The key observation is that the squared covariant derivative is the operator defining the eigenfunctions of the system and implies Landau level quantization quite generically. Finally Galilean boost symmetry validates our formula in Eq. (\ref{generaltopologicalterm}) based on Eq. (\ref{freetheoryrelations}). From there we were able to argue that the Hall viscosity as presented in Eq. (\ref{generaltopologicalterm}) is quantized and should be protected against infinitesimal variations of the theory which respect the assumed symmetries. We then proceeded to derive the non-renormalization of the Hall viscosity under Coulomb interactions. Switching on Coulomb interactions perturbatively implies a finite variation which therefore required a separate treatment. In comparison to the case of integer quantum Hall phases the fractional Jain phases acquire an additional contribution to their Hall viscosity due to the topological spin of composite fermions relative to electrons given by Eqs. (\ref{shifthallviscosity}) and (\ref{shifttopologicalspin}), respectively.\par
In a final step we may now combine the topological expressions for the Hall conductivity in Eq. (\ref{calM4}) and the Hall viscosity in order to provide a Green function representation of the average orbital spin quantum number $\bar{s}$ or equivalently the Wen-Zee shift $\mathcal{S}$ by
\begin{align}
\mathcal{S}=2\bar{s}=4\frac{\mathcal{N}_{\eta_H}+\Delta^{s_{top}}\mathcal{N}_{\eta_H}}{\mathcal{N}_{\sigma_H}}.
\end{align}
This expression comprises the cases for both integer ($s_{top}=0$) and fractional Jain quantum Hall phases ($s_{top}\neq 0$). Employing the geometric and algebraic formalism of Haldane we may be even more precise. Introducing the Landau orbit spin $s_{Lo}$ as well as the guiding center spin $s_{gc}$ we find
\begin{align}
\bar{s}=s_{Lo}+s_{gc},\,\,\,\,s_{Lo}=2\mathcal{N}_{\eta_H}/\mathcal{N}_{\sigma_H},\,\,\,\,s_{gc}=2\Delta^{s_{top}}\mathcal{N}_{\eta_H}/\mathcal{N}_{\sigma_H}.
\end{align}
While the Landau orbit spin originates from non-interacting electron cyclotron motion in the external background magnetic field, the guiding center spin comprises information of strongly correlated clusters of electrons.\par
Finally note that we performed metric tensor calculations throughout in the spirit of non-relativistic differential geometry using the metric tensor as a tool to derive the stress and the Hall viscosity by perturbations around a flat space background. Haldane introduced a dynamical emergent metric relevant for fractional quantum Hall fluids composed of strongly correlated electrons to investigate the Hall viscosity in these phases.

\section{Conclusions}
\label{SectConclusions}

In this work, we introduced a novel Green function expression for the Hall viscosity valid for Galilean invariant fluids within Wigner-Weyl calculus and valid as long as the current density vanishes. Topological quantization of the Hall viscosity for a two dimensional electron gas in a strong magnetic field is known to occur in the case of both translational and rotational invariance modulo the spatial dependence of the gauge field giving rise to the background magnetic field. We were able to proof the topological invariance of the novel Green function expression by variation only at leading order in the filling factor. We found the Green function expression to be identical for Lorentz invariant theories and specifically considered it within the low energy emergent Dirac fermions in graphene \cite{selch2026graphene}. It seems that a major application of our Green function expression for the Hall viscosity is in valleytronics, as its explicit calculation appears to regularize divergences found in a straightforward but algebraically involved Kubo formula analysis of the individual valley Hall viscosities of gapped graphene \cite{sherafati2019hall}. This enables for the first time an analysis of the viscous valley Hall effect \cite{selch2026valley} and is a direct extension of studies of the present work and comes with the prospect of algebraic simplicity of Hall viscosity calculations. A relevant counterexample for the validity of our Green function expression is the low energy two component expansion of bilayer graphene which comprises chiral fermions with quadratic dispersion relation. While the Hall viscosity is only faithfully represented to leading order in the filling fraction in our formula, the valley Hall viscosity seems to be extractable exactly. These findings motivate further study of our findings. In order to check the consistency of our formula for the Hall viscosity, we performed an explicit calculation of its Green function expression for non-interacting electrons as well as composite fermion Jain states in the mean field approximation for $p$ filled (effective) Landau levels and compared it with existing literature. The calculation follows steps analogous to those of our earlier work on the Hall conductivity and yields consistent results.\par
The topological spin characteristic of composite fermions implies that their Wen-Zee shift and therefore their Hall viscosity is shifted as compared to the case of electrons. Extending considerations of \cite{cho2014geometry} we further corroborate the equivalence of the mean field theory model of Lopez and Fradkin to the standard model of electrons, both subject to Coulomb interactions, even in curved spaces. This means that a Jain fractional quantum Hall phase for electrons is equivalent to an integer quantum Hall phase for composite fermions with additional topological spin. Both give rise to the same Hall conductivity and, due to the topological spin, to the same Hall viscosity.\par
The proof of perturbative non-renormalization of the topological expression for the Hall viscosity in the presence of Coulomb interactions builds on our previous work where we reduced the perturbative non-renormalization of the Hall conductivity from a current density analysis in the laboratory frame to that of the density in the Hall fluid frame. By exploiting Galilean invariance we extended these ideas to stresses and momentum densities as relevant for the Hall viscosity.\par
A future objective may be the analysis of the topological character of the response implied by the thermal Hall effect within Wigner-Weyl calculus. We foresee the existence of another Green function expression tailored to the thermal Hall response as derived from a geometric gauge field. Wigner-Weyl calculus is furthermore ideally suited for investigations of the change of the Hall viscosity due to weak spatial inhomogeneities or anisotropies such as produced by topological defects or a phase transition to, e. g., a smectic or nematic phase \cite{you2014theory}. We consider an elaborate discussion of explicit examples of symmetry breaking configurations and their effect on our Green function expression of the Hall viscosity as another worthwhile future study which we consider beyond the scope of this work. In addition, our Green function expression will loose faithfulness and no longer represent the Hall viscosity exactly.

\section*{Acknowledgments}
I thank M. A. Zubkov for discussions and useful remarks on the manuscript.

\vspace{2cm}

\Large{\centerline{\bf{Appendix}}}
\normalsize
\appendix
\numberwithin{equation}{section}

\section{The effective action and hydrodynamics for field theories with non-relativistic diffeomorphism invariance derived from the relativistic theory}
\label{appendix0}

We justify now our definition of the stress tensor given in the main text by considering non-relativistic field theory as the limit of relativistic high energy field theory physics.
We consider field theories which couple to an external spatial metric and an external Abelian gauge field and may be derived as the non-relativistic limit of a high energy theory. The relativistic extension is assumed to be both diffeomorphism invariant with the associated conserved energy momentum tensor $T_{\mu\nu}$ and gauge invariant with associated conserved electric current $j^{\mu}$. Our primary application is the relativistic extension of the model by Lopez and Fradkin derived in \cite{selch2025hall} whose non-relativistic limit yields the known field theory model suggested originally by Lopez and Fradkin \cite{Fradkin1991chern}. The high and low energy regimes are specified by $E\gg m$ and $E\ll m$, respectively, with $m$ being the mass of the relevant quasiparticles (we assume the existence of one species for simplicity) and $E$ a relevant energy scale.\par
The Minkowski spacetime partition function $\mathcal{Z}$ and the effective action $\mathcal{W}$, which depend on the external gauge fields, are related by
\begin{align}
\mathcal{Z}[A_{\mu},g_{\mu\nu}]=e^{i\mathcal{W}[A_{\mu},g_{\mu\nu}]}
\end{align}
with metric tensor $g_{\mu\nu}$ and gauge field $A_{\mu}$ and give rise to the conserved currents via the relations ($g=det(g_{\mu\nu})$)
\begin{align}
T_{\mu\nu}=\frac{2i}{\sqrt{-g}}\frac{1}{\mathcal{Z}}\frac{\delta \mathcal{Z}}{\delta g^{\mu\nu}}=-\frac{2}{\sqrt{-g}}\frac{\delta \mathcal{W}}{\delta g^{\mu\nu}},\,\,\,\, j^{\mu}=i\frac{1}{\mathcal{Z}}\frac{\delta \mathcal{Z}}{\delta A_{\mu}}=-\frac{\delta \mathcal{W}}{\delta A_{\mu}}.
\end{align}
Diffeomorphisms and gauge transformations leave the effective action invariant. The linear order variational expansion of
\begin{align}
\mathcal{W}[A_{\mu}+\delta A_{\mu},g_{\mu\nu}+\delta g_{\mu\nu}]=\mathcal{W}[A_{\mu},g_{\mu\nu}]
\end{align}
implies the Minkowski spacetime ($g_{\mu\nu}=\eta_{\mu\nu}$) hydrodynamic equations
\begin{align}
\partial_{\mu}T^{\mu\nu}=0,\,\,\,\, \partial_{\mu}j^{\mu}=0\,\,\,\,(x^{\mu}=(ct,\bold{x})).
\end{align}
We may follow these relations from the relativistic theory at high energies to its low energy regime \cite{SonWingate2006}. We will assume a vanishing relativistic Abelian gauge field, as such a gauge field will be regenerated from the temporal components of the metric. In the non-relativistic limit we take $c\to \infty$ with $c$ being the (reinstated) velocity of light in vacuum and employ the (inverse) metric expansions
\begin{align}
g_{\mu\nu}=\begin{pmatrix}
-1-\frac{2A_0}{mc^2} & -\frac{A_i}{mc} \\
-\frac{A_i}{mc} & g_{ij}
\end{pmatrix},\,\,\,\,
g^{\mu\nu}=\begin{pmatrix}
-1+\frac{2A_0}{mc^2}+\frac{A_ig^{ij}A_j}{m^2c^2}+O(\frac{1}{c^4}) & -\frac{g^{ij}A_j}{mc}+O(\frac{1}{c^3}) \\
-\frac{g^{ij}A_j}{mc}+O(\frac{1}{c^3}) & g^{ij}+O(\frac{1}{c^2})
\end{pmatrix}.
\end{align}
The invariance under diffeomorphisms at high energies manifests itself in the form of the so-called non-relativistic diffeomorphism invariance or gauged Galilean invariance at low energies. Proceeding as in the relativistic case we obtain the hydrodynamic equations 
\begin{align}
\partial_0T^0_{\,\,\,k}+\partial_iT^i_{\,\,\,k}=0,\,\,\,\,\partial_0\rho +\partial_ij^i=0
\end{align}
where
\begin{align}
\nonumber \rho =&i\frac{1}{\mathcal{Z}}\frac{\delta \mathcal{Z}}{\delta A_0}=-\frac{\delta \mathcal{W}}{\delta A_0},\,\,\,\,j^i=i\frac{1}{\mathcal{Z}}\frac{\delta \mathcal{Z}}{\delta A_i}=-\frac{\delta \mathcal{W}}{\delta A_i},\\
\nonumber T^0_{\,\,\,k}=&img_{ik}\frac{1}{\mathcal{Z}}\frac{\delta \mathcal{Z}}{\delta A_i}-iA_k\frac{1}{\mathcal{Z}}\frac{\delta \mathcal{Z}}{\delta A_0}=-mg_{ik}\frac{\delta \mathcal{W}}{\delta A_i}+A_k\frac{\delta \mathcal{W}}{\delta A_0},\\
T^i_{\,\,\,k}=&2ig_{kj}\frac{1}{\mathcal{Z}}\frac{\delta \mathcal{Z}}{\delta g_{ij}}-i\delta^i_k\log\mathcal{Z}+iA_k\frac{1}{\mathcal{Z}}\frac{\delta \mathcal{Z}}{\delta A_i}=-2g_{kj}\frac{\delta \mathcal{W}}{\delta g_{ij}}+\delta^i_k\mathcal{W}-A_k\frac{\delta \mathcal{W}}{\delta A_i}.
\label{chargesandcurrents}
\end{align}
In order to extract the Hall viscosity from the spatial stress tensor $T^i_{\,\,\,k}$ we only need to consider the traceless part. In the case of rotational invariance, which implies $T^0_{\,\,\,i}=j^i=0$, we may therefore assume with $g_{ij}=\delta_{ij}+\delta g_{ij}$ and $\delta^{ij}\delta g_{ij}=0$ to leading order that
\begin{align}
T^i_{\,\,\,k}=2ig_{kj}\frac{1}{\mathcal{Z}}\frac{\delta\mathcal{Z}}{\delta g_{ij}}\approx 2i\delta_{kj}\frac{1}{\mathcal{Z}}\frac{\delta\mathcal{Z}}{\delta g_{ij}}\,\,\Leftrightarrow\,\,T_{ij}=2i\frac{1}{\mathcal{Z}}\frac{\delta\mathcal{Z}}{\delta g^{ij}}.
\end{align}
This form has been used in the main text. Notice also the relation $T^0_{\,\,\,i}=mj^i$ implied by (local) Galilean invariance.

\section{Proof of the equivalence of the curved space generalized Lopez and Fradkin composite fermion model and electrons in curved space subject to Coulomb interactions}
\label{appendixA}

We consider the action 
\begin{align}
S[J,\omega ,\mathcal{A}]=S_{CS}[\mathcal{A}]+S_{source}[J,\omega ,\mathcal{A}]
\end{align} 
with constituent terms
\begin{align}
\nonumber &S_{CS}[\mathcal{A}]=\frac{ \theta}{4} \int d^3x \epsilon^{\mu\nu\rho} \mathcal{A}_{\mu}\mathcal{F}_{\nu\rho},\\
&S_{source}[J,\omega ,\mathcal{A}]=-\int d^3x J^{\mu}(\mathcal{A}_{\mu}+s\omega_{\mu})
\end{align}
on a manifold of the form $\Sigma^2\times\mathbb{R}$ with spatial component $\Sigma^2$ and time line $\mathbb{R}$ whereby $\theta =(2\pi 2s)^{-1}$ together with current conservation $\nabla_{\mu}J^{\mu}=0$. We intend to prove that even in curved space the standard model of electrons interacting via Coulomb forces is equivalent to the composite fermion theory model of Lopez and Fradkin. In curved space, the spatial metric $g_{ij}$ depends on coordinates and gives rise to a non-vanishing Abelian spin connection $\omega_{\mu}$ in the non-relativistic limit.\par
The proof of equivalence in flat spacetime may be found in quantum mechanical and field theoretical path integral formulations in appendices of \cite{Fradkin1991chern,selch2025hall}, respectively, and has essentially been extended in the presence of curvature in section II. of \cite{cho2014geometry} and references mentioned there. We intend to refine the arguments presented therein. According to \cite{Fradkin1991chern} we may perform the Gaussian path integral manipulations
\begin{align}
\nonumber e^{iS_{eff}[J,\omega ]}=&\int D\mathcal{A}e^{iS[J,\omega ,\mathcal{A}]}=\int D\mathcal{A}e^{i\int d^3x \Big[\Big(\int d^3y\frac{1}{2}\mathcal{A}_{\mu}(x)K^{\mu\nu}(x,y)\mathcal{A}_{\nu}(y)\Big)-J^{\mu}(x)(\mathcal{A}_{\mu}(x)+s\omega_{\mu}(x))\Big]}\\
=&\mathcal{N}e^{i\int d^3x \Big[\Big(\int d^3y \frac{1}{2}J^{\mu}(x)(K^{-1})_{\mu\nu}(x,y)J^{\nu}(y)\Big)-J^{\mu}(x)s\omega_{\mu}(x)\Big]}
\end{align}
with
\begin{align}
\nonumber &K^{\mu\nu}(x,y)=\theta\delta^3(x-y)\epsilon^{\mu\nu\rho}\nabla^y_{\rho},\,\,\,\, K^{-1}_{\mu\nu}(x,y)=\frac{1}{\theta}G_{\mu\nu}(x,y),\\
&G_{\mu\nu}(x,y)=-\frac{1}{4\pi}\epsilon_{\mu\nu\rho}\nabla_x^{\rho}\frac{1}{|x-y|^2}
\end{align}
and normalization factor $\mathcal{N}$. If $J^{\mu}$ originates from point-like particles with charges $q_i$ in units of the elementary charge traversing relatively disjoint, closed trajectories $C_i$ such that
\begin{align}
J^{\mu}(x)=\sum_i q_i \oint_{C_i}d\tau T_i^{\mu}(\tau )\delta^3(x-X_i(\tau )) 
\end{align}
with curve tangent vector field $T^{\mu}_i=\frac{dx_i^{\mu}}{d\tau}$ then the effective action $S_{eff}[J,\omega ]$ evaluates to 
\begin{align}
S_{eff}[J,\omega ]=&2\pi s\Big[\sum_{i\neq j}q_iq_jLk(C_i,C_j)+\sum_iq_i^2Sl(C_i)\Big]-s\sum_i\int_{C_i}dx_i^{\mu}\omega_{\mu}.
\end{align}
The topological Gauss linking number $Lk$ and self-linking number $Sl$ are explained in some more detail subsequently. On $\mathbb{R}^3$ the Gauss linking number of two disjoint curves is defined by
\begin{align}
Lk(C_i,C_j)=&\int_{C_i\times C_j}\omega (x_i-x_j),\,\,\,\,\omega (x)=\frac{1}{8\pi}\frac{\epsilon_{\mu\nu\rho}x^{\mu}dx^{\nu}dx^{\rho}}{|x|^3}.
\end{align}
More generally on three dimensional manifolds, the Gauss linking number is topologically quantized and restricted to integer values. The self-linking number is more subtle, as it has singularities at coincident points. To define the self-linking number $Sl(C)$ for a closed curve $C$ properly one must introduce a regularization scheme. A gauge invariant choice is to displace $C$ infinitesimally in a normal direction and evaluate the mutual linking number of C and its displaced version. Alternatively one may think of thickening the curve to ribbon and compute the linking of the curves traversed by the ribbon edges. The resulting scheme ambiguity is known as the framing anomaly within Chern-Simons theory. More formally the framing of the Wilson line operator on an oriented curve is a choice of nowhere zero section $\sigma^{\mu}$ of the normal bundle of the curve up to homotopy. Within Chern-Simons theory this prescription is used to evaluate Wilson loop observables which are UV singular on coincident points. According to the Calugareanu-White-Fuller theorem the so defined, regularized self-linking number fulfills the equality
\begin{align}
Sl(C,\sigma )=Tw(C,\sigma )+Wr(C)
\end{align}
with the twist $Tw$ and the writhe $Wr$ of the curve $C$. While the twist depends on the framing $\sigma$, the writhe does not. They are defined in the following way
\begin{align}
Tw(C,\sigma )=&\frac{1}{2\pi}\int_C d\tau\epsilon_{\mu\nu\rho}T^{\mu}\sigma^{\nu}\frac{d\sigma ^{\rho}}{d\tau},\\
Wr(C)=&PV\int_{C\times C}\omega (x-x^{\prime}),\,\,\,\,\omega (x)=\frac{1}{8\pi}\frac{\epsilon_{\mu\nu\rho}x^{\mu}dx^{\nu}dx^{\rho}}{|x|^3}.
\end{align}
Neither of these terms is topological on its own. $PV$ stands for principal value which excludes a symmetric $\epsilon$-tube around the singular diagonal of the double integral. Different framings differ by an integer number of extra twists $Tw\to Tw+m,\,m\in\mathbb{Z}$ and manifest the framing anomaly. The decomposition has been proven in $\mathbb{R}^3$ with proper extensions to $S^3$ (3-sphere) and $H^3$ (hyperbolic space) \cite{deturck2010linking}.\par
We now postulate the Calugareanu-White-Fuller theorem, as applied to fermionic particles, to hold in curved spaces such that both twist and writhe of the regularized self-linking curve may be split into flat space and geometric contributions. The non-relativistic contribution to the geometric part of the writhe of a curve $C$ in a general curved space is to be defined by an Abelian spin connection via
\begin{align}
Wr^{geo}(C)=-\frac{1}{2\pi}\int_Cdx^{\mu}\omega_{\mu}.
\label{geowrithe}
\end{align}
and will be motivated shortly. This definition depends only on the curve and the local geometry around it which is why it exists independent of any regularization of the self-linking. This term is exactly the term we added for fractional quantum Hall states in the Jain hierarchy and which multiplies the topological orbital spin. These considerations imply for the effective action we aim to evaluate (with unit charge composite fermions)
\begin{align}
\nonumber S_{eff}[J,\omega ]=&2\pi sn+2\pi s Tw^{flat}(C_i,\sigma )+2\pi s Tw^{geo}(C_i,\sigma )+2\pi sWr^{flat}(C_i)+2\pi sWr^{geo}(C_i)\\
\nonumber =&2\pi sn+s\int_{C_i} d\tau\epsilon_{\mu\nu\rho}T^{\mu}\sigma^{\nu}\frac{d\sigma ^{\rho}}{d\tau}+2\pi sWr^{flat}(C_i)-s\int_{C_i}dx^{\mu}\omega_{\mu},\,\,\,\,n\in\mathbb{Z}.
\end{align}
It has been shown in detail in section II. of \cite{cho2014geometry} (notice the mistake in Eq. (2.2) therein where the writhe in the intermediate term should be replaced by the linking) that in curved spacetime (with Chern-Simons level $k=2\pi\theta$) in the non-relativistic limit
\begin{align}
\int_C d\tau\epsilon_{\mu\nu\rho}T^{\mu}\sigma^{\nu}\frac{d\sigma ^{\rho}}{d\tau}=Tw^{flat}(C,\sigma )+\int_Cdx^{\mu}\omega_{\mu}
\end{align}
justifying Eq. (\ref{geowrithe}) and finally leading to
\begin{align}
S_{eff}[J,\omega ]=2\pi sn+2\pi sm\,\,\,\,n,m\in\mathbb{Z}.
\end{align}
Thereby both the topological character of the effective action and the presence of the framing anomaly are ensured. In conclusion the curved space generalized Lopez and Fradkin composite fermion model has to be supplemented by a coupling of the fermionic current $J^{\mu}$ to the Abelian spin connection $\omega_{\mu}$ with the strength determined by the composite fermion topological spin $s$ modulo regularization dependent modifications. This generalizes the equivalence of the mean field theory model of Fradkin and Lopez to the standard model of electrons, both in the presence of Coulomb interactions, to curved spaces with the mentioned reservation.

\section{Expression for $\mathcal{N}_{\eta_H}$ through the effective Hamiltonian}
\label{appendixC}

We proceed now to calculate $\mathcal{N}_{\eta_H}$ explicitly within the mean field approximation neglecting topological spin contributions and thus follow the steps in the calculation of $\mathcal{N}_{\sigma_H}$ as presented in Appendix D of \cite{selch2025hall} and adapted to $\mathcal{N}_{\eta_H}$. For definiteness we choose the Landau gauge in which the vector potential $\bold{A}$ is given by
\begin{align}
A_x = 0, \quad A_y = B x\,. 
\end{align}
The effect of interactions is included through the effective magnetic field $B_{eff} = B - \rho_0 / \theta$ \cite{Fradkin1991chern}. We will start from the expression in Eq. (\ref{generaltopologicalterm})
\begin{align}  \label{N-2}
\nonumber \mathcal{N}_{\eta_H}=&{\frac{T}{16\pi^2 A}}\int d^3p d^3x \frac{\delta^{lm}}{B_{eff}}\epsilon_{ij}\Tr\Big[ (G_{eff})_W\star \partial_{p_3} (Q_{eff})_W\star (G_{eff})_W\star\partial_{p_i} (Q_{eff})_W\star (D_W)_l\\
&\star (G_{eff})_W \star \partial_{p_j} (Q_{eff})_W\star (D_W)_m \Big]. 
\end{align}
In Wigner representation, the quantity $\partial_{p_i} (Q_{eff})_W$ is calculated as follows
\begin{align}
\nonumber \frac{\partial}{\partial p^i}(Q_{eff})_W(x,p)=&\frac{\partial}{\partial p^i}\int \frac{d^3P}{(2\pi )^3}~e^{iPx} \tilde{Q}_{eff}\Big(p+\frac{P}{2},p-\frac{P}{2}\Big)\\
=&\int \frac{d^3P}{(2\pi )^3}~e^{iPx}\Big(\frac{\partial}{\partial K_1^i }+\frac{\partial}{%
\partial {K_2^i} }) \tilde{Q}_{eff}(K_1,K_2)\Big|_{K_1=p+\frac{P}{2}}^{K_2=p-\frac{P%
}{2}}
\end{align}
where we used the chain rule of differentiation to get 
\begin{align}
\frac{\partial}{\partial {p^i} } =\frac{\partial K_1^j}{\partial {p^i} }%
\frac{\partial}{\partial {K_1^j} }+\frac{\partial K_2^j}{\partial {p^i} }%
\frac{\partial}{\partial {K_2^j} } =\frac{\partial}{\partial {K_1^i} }+\frac{%
\partial}{\partial {K_2^i} }.
\end{align}
If we denote 
\begin{align}
\nonumber &\tilde{Q}_{eff,i}\Big(p+\frac{P}{2},p-\frac{P}{2}\Big)=\Big( \frac{\partial}{\partial {K_1^i}} +\frac{\partial}{\partial{K_2^i}}\Big) 
\tilde{Q}_{eff}(K_1,K_2)\Big|_{K_1=p+P/2}^{K_2=p- P/2}
\end{align}
then $\partial_{p_i}(Q_{eff})_W$ can be written in short-hand form as $\partial_{p_i}(Q_{eff})_W=(Q_{eff,i})_W$. Using the associativity of the Moyal star product Eq. (\ref{N-2}) is written as 
\begin{align}
\nonumber &(G_{eff})_W \star \partial_{p_3} (Q_{eff})_W \star (G_{eff})_W\star\partial_{p_i} (Q_{eff})_W\star (D_W)_l \star (G_{eff})_W \star \partial_{p_j} (Q_{eff})_W\star (D_W)_m\\
&=\int \frac{d^3P}{(2\pi )^3}~e^{iPx}(\tilde{G}_{eff}\tilde{Q}_{eff,3}\tilde{G}_{eff}\tilde{Q}_{eff,j}\tilde{D}_l\tilde{G}_{eff}\tilde{Q}_{eff,k}\tilde{D}_m)
\Big(p+\frac{P}{2},p-\frac{P}{2}\Big)
\end{align}
where 
\begin{align}
&(\tilde{G}_{eff}\tilde{Q}_{eff,3}\tilde{G}_{eff}\tilde{Q}_{eff,i}\tilde{D}_l\tilde{G}_{eff}\tilde{Q}_{eff,j}\tilde{D}_m)\Big(p+\frac{P}{2},p-\frac{P}{2}\Big) = \nonumber \\
&  \int d^3p^{(2)}d^3p^{(3)}d^3p^{(4)}d^3p^{(5)}d^3p^{(6)}d^3p^{(7)}\delta^3\Big(p^{(7)}-p+\frac{P}{2}\Big)\Big[ \tilde{G}_{eff}(p + \frac{P}{2},p^{(2)})\Big( \lbrack\partial_{p^{(2)}_3} + \partial_{p^{(3)}_3}] \tilde{Q}_{eff}(p^{(2)},p^{(3)})\Big)\cdot \nonumber \\
&\tilde{G}_{eff}(p^{(3)},p^{(4)})\Big( \lbrack \partial_{p^{(4)}_i} +\partial_{p^{(5)}_i}] \tilde{Q}_{eff}(p^{(4)},p^{(5)})\tilde{D}_l\Big)\tilde{G}_{eff}(p^{(5)},p^{(6)})\Big( \lbrack \partial_{p^{(6)}_j} +\partial_{p^{(7)}_j}] \tilde{Q}_{eff}(p^{(6)},p^{(7)})\tilde{D}_m\Big) \Big].  
\label{GQ0GQiGQj}
\end{align}
Substituting the above expressions back into Eq. (\ref{N-2}) and integrating over $x$ and $P$ leads to ($T$ vanishes due to imaginary time independence of the propagator (and its inverse) and integration over $d\tau$)
\begin{align}
\mathcal{N}_{\eta_H}=&{\frac{1}{16\pi^2A}} \int d^3p^{(1)}d^3p^{(2)}d^3p^{(3)}d^3p^{(4)}d^3p^{(5)}
d^3p^{(6)}\frac{\delta^{lm}\epsilon_{ij}}{B_{eff}}\Tr\Big[\tilde{G}_{eff}(p^{(1)} ,p^{(2)})\cdot\nonumber\\
&\Big( \lbrack\partial_{p^{(2)}_3} + \partial_{p^{(3)}_3}] \tilde{Q}_{eff}(p^{(2)},p^{(3)})\Big)\tilde{G}_{eff}(p^{(3)},p^{(4)})\Big( \lbrack \partial_{p^{(4)}_i} +\partial_{p^{(5)}_i}] \tilde{Q}_{eff}(p^{(4)},p^{(5)})\tilde{D}_l\Big)\cdot\nonumber\\
&\tilde{G}_{eff}(p^{(5)},p^{(6)})\Big( \lbrack \partial_{p^{(6)}_j} +\partial_{p^{(1)}_j}] \tilde{Q}_{eff}(p^{(6)},p^{(1)})\tilde{D}_m\Big)\Big].
\label{GQ0GQiGQj-2}
\end{align}
The momentum space function $\tilde{Q}_{eff}(p^{(i)},p^{(j)})$ has the representation 
\begin{align}
\tilde{Q}_{eff}(p^{(i)},p^{(j)}) = \langle p^{(i)}| \hat{Q}_{eff} | p^{(j)}\rangle =\Big(  i\omega^{(i)}\delta^{2} (p^{(i)} - p^{(j)}) - \langle p^{(i)} |
\hat{H}_{eff} | p^{(j)} \rangle \Big) \delta (\omega^{(i)}-\omega^{(j)})  \label{Q1}
\end{align}
where $p = (p_3,p_1,p_2) = (\omega, \mathbf{p})$. The corresponding Green function can be calculated as 
\begin{align}
\tilde{G}_{eff}(p^{(i)},p^{(j)})= \sum_{n} \frac{1}{i\omega^{(i)} - \mathcal{E}_n}
\langle {p}^{(i)}| n \rangle \langle n | p^{(j)}\rangle
\delta(\omega^{(i)}-\omega^{(j)})\,. 
\end{align}
The numbers $\mathcal{E}_n$ represent the energy eigenvalues of $\hat{H}_{eff}$ shifted by the chemical potential $\mu$. From Eq. (\ref{Q1}) it may be concluded that 
\begin{align}
\nonumber \partial_{p^{(i)}_3} \tilde{Q}_{eff} (p^{(i)},p^{(i+1)})=& i\delta^{2} (p^{(i)}- p^{(i+1)})\delta (\omega^{(i)}-\omega^{(i+1)})(\partial_{p^{(i)}_j} + \partial_{p^{(i+1)}_j}) \tilde{Q}_{eff} (p^{(i)},p^{(i+1)})  \\
\nonumber =& -(\partial_{p^{(i)}_j} + \partial_{p^{(i+1)}_j}) \langle p^{(i)} |
\hat{H}_{eff} | p^{(j)} \rangle \delta(\omega^{(i)}-\omega^{(i+1)})  \\
\nonumber =& -i \langle p^{(i)} | \hat{H}_{eff} \hat x_j - \hat x_j \hat{H}_{eff} | p^{(i+1)}\rangle \delta(\omega^{(i)}-\omega^{(i+1)}) \\
=& -i \langle p^{(i)} | [\hat{H}_{eff}, x_j]| p^{(i+1)}\rangle \delta(\omega^{(i)}-\omega^{(i+1)}).
\label{pro-Q}
\end{align}
We can easily integrate out all intermediate $\omega^{(i)}$'s by means of the delta functions to remain with a single $\omega$-integration and then proceed to integrate
out $(p^{(2)},p^{(3)})$ in Eq. (\ref{GQ0GQiGQj-2}) after substituting the expression of Eq. (\ref
{pro-Q}) and using the completeness relation $\hat{1}=\int d^3p_i|p_i \rangle \langle p_i |$. After such operations Eq. (\ref{GQ0GQiGQj-2}) reduces to 
\begin{align}
\nonumber \mathcal{N}_{\eta_H}=&\frac{i}{16\pi^2A}\,\sum_{n,k} \int d\omega d^2p^{(1)}d^2p^{(2)}d^2p^{(3)}d^2p^{(4)}\frac{\delta^{lm}}{B_{eff}} \epsilon_{ij}\Tr\,\Big[ \frac{1}{(i\omega^{} - \mathcal{E}_n)^2} \langle {\ p}^{(1)}| n \rangle \langle n | {\ p}^{(2)}\rangle\cdot\\ 
&\langle {p}^{(2)}| [\hat{H}_{eff}, \hat x_i]\tilde{D}_l | {\ p}^{(3)}\rangle\frac{1}{(i\omega^{} - \mathcal{E}_k)}\langle {p}^{(3)}| k \rangle \langle k | {\ p}^{(4)}\rangle \langle {\ p}^{(4)}| [\hat{H}_{eff}, \hat x_j] \tilde{D}_m| {p}^{(1)}\rangle \Big]
\end{align}
where we redefine the momentum variables left as $p^{(1)}, p^{(2)},p^{(3)}, p^{(4)}$. Further integrating out the momenta $p^{(i)}$ we have 
\begin{align}  
\label{N-3}
\nonumber \mathcal{N}_{\eta_H} =&\frac{i}{16\pi^2A}\,\sum_{n,k} \int \,d \omega d^2p\frac{\delta^{lm}}{B_{eff}}
\epsilon_{ij}\,\Big[ \frac{1}{(i\omega^{} - \mathcal{E}_n)^2}\langle n| [\hat{H}_{eff}, {\hat x}_i]\tilde{D}_l | k \rangle \frac{1}{(i\omega^{} - \mathcal{E}_k)}
\langle k | [\hat{H}_{eff}, {\hat x}_j]\tilde{D}_m | n \rangle \Big]\\
=& -\frac{i}{8\pi A}\,\sum_{n,k}\frac{\delta^{lm}}{B_{eff}} \epsilon_{ij}\, \langle n | [\hat{
H}_{eff}, \hat x_i]\tilde{D}_l | k \rangle \langle k | [\hat{H}_{eff}, \hat x_j] \tilde{D}_n| n
\rangle\frac{ (\Theta(- \mathcal{E}_n) \Theta(\mathcal{E}_k) - \Theta(- 
\mathcal{E}_k) \Theta(\mathcal{E}_n))}{(\mathcal{E}_k - \mathcal{E}_n)^2}. 
\end{align}
In order to obtain the second equality in Eq. (\ref{N-3}) we have employed the residue theorem and compactified the resulting expression using Heaviside step functions $\Theta$.\par
Now, we will define new variables $(\xi_i, X_i)$ as follows 
\begin{align}
\hat{x}_1 = -\frac{\hat{p}_y - B_{eff} x}{B_{eff}} + \hat{X}_1 = \hat{\xi}_1+ \hat{X}_1,\,\,\,\,\hat{x}_2 = \frac{\hat{p}_x}{B_{eff}} + \hat{X}_2= \hat{\xi}_2 + \hat{X}_2,\,\,\,\,-iD_l=B_{eff}\epsilon_{lm}\hat{\xi}_m. 
\end{align}
The commutation relations implied by the above set of variables are 
\begin{align}
[\hat{\xi}_{i},\hat{\xi}_{j}] = \frac{i}{B_{eff}}\epsilon_{ij}, \,\,\,\, [\hat{X}_{i},\hat{X}_{j}] = - \frac{i}{ B_{eff}}\epsilon_{ij},\,\,\,\,[\hat{H}_{eff}, \hat{X}_{i}] = 0 \,\,\,\,\Big(\hat{H}_{eff}=\frac{B_{eff}^2}{2m}(\hat{\xi}_1^2+\hat{\xi}_2^2)\Big)
\end{align}
where $i$ labels the first and second set of variables, respectively. In terms of the $\hat{\xi}_i$'s, Eq. (\ref{N-3}) can be written as 
\begin{align}
A\mathcal{N}_{\eta_H} =\frac{iB_{eff}^2}{8\pi} \sum_{n,k} \epsilon_{ij}\frac{\delta^{lm}}{B_{eff}} \,\Big[ \langle n| [\hat{H}
_{eff}, {\hat \xi}_i] \hat{\xi}_l| k \rangle\langle k | [\hat{H}_{eff}, {\hat \xi}_j]\hat{\xi}_m | n
\rangle \Big] \frac{( \Theta(-\mathcal{E}_n)\Theta(\mathcal{E}_k) - \Theta(-
\mathcal{E}_k)\Theta(\mathcal{E}_n))}{(\mathcal{E}_k - \mathcal{E}_n)^2}.
\end{align}
The two terms involving Heaviside step functions entering with opposite signs give identical contributions to $A\mathcal{N}_{\eta_H}$ which may be checked easily using Eqs. (\ref{matrixelements}) and (\ref{matrixelements2}) below. We may then insert two further resolutions of the identity in terms of energy eigenstates to write
\begin{align}
\nonumber A\mathcal{N}_{\eta_H} =&\frac{iB_{eff}^2}{4\pi} \sum_{n,k,p,r} \epsilon_{ij}\frac{\delta^{lm}}{B_{eff}} \,\Big[ \langle n| [\hat{H}
_{eff}, {\hat \xi}_i]|q\rangle \langle q| \hat{\xi}_l| k \rangle\langle k | [\hat{H}_{eff}, {\hat \xi}_j]|r\rangle \langle r|\hat{\xi}_m | n
\rangle \Big] \frac{\Theta(-\mathcal{E}_n)\Theta(\mathcal{E}_k)}{(\mathcal{E}_k - \mathcal{E}_n)^2}\\
=&\frac{iB_{eff}^2}{4\pi} \sum_{n,k,p,r} \epsilon_{ij}\frac{\delta^{lm}}{B_{eff}} \,\Big[ \langle n|{\hat \xi}_i|q\rangle \langle q| \hat{\xi}_l| k \rangle \langle k | {\hat \xi}_j|r\rangle \langle r|\hat{\xi}_m | n
\rangle \Big]\frac{(\mathcal{E}_n-\mathcal{E}_q)\cdot (\mathcal{E}_k-\mathcal{E}_r)}{(\mathcal{E}_k - \mathcal{E}_n)^2}\Theta(-\mathcal{E}_n)\Theta(\mathcal{E}_k).
\label{hallviscosityintermediate}
\end{align}
The indices $n$, $k$, $q$, $r$ have a discrete index (dis) labeling Landau levels and a continuous index (con) labeling intra Landau level states. We may now employ the relations for matrix elements of the effective Landau level Hamiltonian assuming for simplicity that $A=L^2$ is square shaped
\begin{align}
&\langle n|\hat{\xi}_1|k\rangle =\frac{l_{Beff}}{L}\delta (n_{con}-k_{con})\Big(\sqrt{\frac{n_{dis}}{2}}\delta_{n_{dis},k_{dis}+1}+\sqrt{\frac{n_{dis}+1}{2}}\delta_{n_{dis},k_{dis}-1}\Big),\label{matrixelements}\\
&\langle n|\hat{\xi}_2|k\rangle =i\frac{l_{Beff}}{L}\delta (n_{con}-k_{con})\Big(\sqrt{\frac{n_{dis}}{2}}\delta_{n_{dis},k_{dis}+1}-\sqrt{\frac{n_{dis}+1}{2}}\delta_{n_{dis},k_{dis}-1}\Big).
\label{matrixelements2}
\end{align}
The quantity $l_{B_{eff}}$ is known as the magnetic length and parametrizes the characteristic extension of electron wavefunctions (see, e. g., \cite{Tong:2016kpv}). We assume that in the effective field theory $p$ effective Landau levels are fully filled resulting in $\mathcal{N}_{\sigma_H}=p$ and thereby an effective IQHE. Furthermore our momentum space scaling and the Landau level degeneracy imply the intra Landau level momentum range
\begin{align}
\Delta p_y=2\pi \frac{L}{l_{B_{eff}}^2},\,\,\,\, l_{B_{eff}}^2=(B_{eff})^{-1}.
\label{momentumrange}
\end{align}
Employing both Eqs. (\ref{matrixelements}), (\ref{matrixelements2}) and (\ref{momentumrange}) within Eq. (\ref{hallviscosityintermediate}) leads to $\mathcal{N}_{\eta_H}=\frac{1}{4}p^2$
for the topological invariant for the Hall viscosity in the effective mean field approximation. Notice that in the course of the calculation every step stays identical if we perform a variation of the Hamiltonian $\hat{H}_{eff}$ to $\hat{H}$ such that
\begin{align}
\hat{H} =\hat{H}_{eff}+\Delta H\Big(\frac{B^2_{eff}}{2m}(\hat{\xi}_1^2+\hat{\xi}_2^2)\Big)
\end{align}
for an arbitrary function $\Delta H$. At the same time our calculation in Eq. (\ref{hallviscosityintermediate}) will only represent the leading term of the Hall viscosity coefficient correctly.\par
While $\mathcal{N}_{\sigma_H}$ counts the number of fully occupied Landau levels in the mean field composite fermion theory within the FQHE or analogously the number of fully occupied Landau levels in the theory of free electrons within the IQHE, $\mathcal{N}_{\eta_H}$ has been found to be proportional to a quantized orbital spin of the emergent quasiparticles within the Hall fluid in the case of rotational invariance (see, e. g., \cite{hoyos2014hall}). The quantized formula is generally expressed as
\begin{align}
\mathcal{N}_{\eta_H}=\frac{1}{2}\bar{s}\nu =\frac{1}{4}\mathcal{S}\nu
\label{wenzeeshifttimesfilling}
\end{align}
with $\bar{s}$ being the average orbital spin per particle, $\mathcal{S}$ is the so-called Wen-Zee shift and $\nu$ is the filling fraction. A filling of $p$ Landau levels by non-interacting electrons corresponds to $2\bar{s}=\mathcal{S}=\nu =p$ and is thereby consistent with our final result. \par
The filling of $p$ effective Landau levels by composite fermions leads to a different result due to the presence of topological spin. The induced shift is proportional to the trace of the effective composite fermion Green function. Evaluated at finite temperature $T$ with chemical potential $\mu$ and subjected to the limit $T\to 0$ a standard calculation with Matsubara frequencies $\omega_m$ and Fermi-Dirac distribution function $n_F(x)=(e^{\beta x}+1)^{-1}$ yields
\begin{align}
\nonumber \lim_{T\to 0}\sum_{n=0}^{\infty}\sum_{\omega_m}\Tr [(G_{eff})^{\omega_m}_n]=&\lim\limits_{T\to 0} \frac{1}{2\pi l_{B_{eff}}^2}\sum_{n=0}^{\infty}T\sum_{\omega_m}\frac{1}{i\omega_m+\mu -\mathcal{E}_n}=\lim\limits_{T\to 0} \frac{1}{2\pi l_{B_{eff}}^2}\sum_{n=0}^{\infty}(-1)n_F(\mathcal{E}_n-\mu )\\
=&\frac{1}{2\pi l_{B_{eff}}^2}\sum_{n=0}^{\infty}(-1)\Theta (\mathcal{E}_n-\mu )=-p\frac{B_{eff}}{2\pi}.
\end{align}
Together with the already stated topological spin induced shift of $\mathcal{N}_{\eta_H}$ in Eq. (\ref{shifttopologicalspin}) we obtain the total expression of the topological invariant in the form of Eq. (\ref{wenzeeshifttimesfilling}) with $\nu =p$ but now $\mathcal{S}=p+2s_{top}=p+2s$. This result is compatible with previous effective field theory calculations at scales well below the ground state energy gap such that fermionic degrees of freedom may be integrated out \cite{cho2014geometry}.

\section{On the genericity of the Green function expression of $\mathcal{N}_{\eta_H}$}
\label{appendixD}

We showed that within linear response theory the Hall viscosity may be expressed via
\begin{align}
\eta_H=2\cdot \frac{1}{4}\epsilon^{ik}\delta^{jl}\frac{\delta \bar{T}_{ij}}{\delta \partial_0g^{kl}}=\frac{1}{2\pi}\mathcal{N}_{\eta_H}B_{eff}
\end{align}
with averaged stress tensor $\bar{T}_{ij}$ and $\mathcal{N}_{\eta_H}$ expressed via Green functions
\begin{align}
\mathcal{N}_{\eta_H}=&-\frac{1}{4\pi^2\beta A}\int d^3x\int d^3p\frac{\epsilon^{ik}\delta^{jl}}{B_{eff}}\Tr\Big[\frac{\delta (Q_{eff})_W(x,p)}{\delta  g^{ij}}\star (G_{eff})_W(x,p)\\
&\star\frac{\delta (Q_{eff})_W(x,p)}{\delta g^{kl}}\star (G_{eff})_W(x,p)\star\frac{\partial (Q_{eff})_W(x,p)}{\partial \omega}\star (G_{eff})_W(x,p)\Big].
\label{NetaH1}
\end{align}
We then applied the identity
\begin{align}
\frac{\delta (Q_{eff})_W}{\delta g^{ij}}=&\frac{i}{2}\frac{\partial (Q_{eff})_W}{\partial p_{(i}}\star (D_W)_{j)}\Big\rvert_{g^{ij}=\delta^{ij}}=\frac{i}{2}(D_W)_{(i}\star\frac{\partial (Q_{eff})_W}{\partial p_{j)}}\Big\rvert_{g^{ij}=\delta^{ij}}
\label{freetheoryrelations2}
\end{align}
with $(D_W)_{\mu}=-ip_{\mu}+iA_{\mu}(x)$, presented in Eq. (\ref{freetheoryrelations}), to show that 
\begin{align}
\nonumber \mathcal{N}_{\eta_H}=&\frac{1}{16\pi^2\beta A}\int d^3x\int d^3p\frac{1}{B_{eff}}\epsilon^{ik}\delta^{jl}\Tr\Big[\frac{\partial (Q_{eff})_W(x,p)}{\partial p_i}\star (D_W)_j\star (G_{eff})_W(x,p)\\
&\star\frac{\partial (Q_{eff})_W(x,p)}{\partial p_k}\star (D_W)_l\star(G_{eff})_W(x,p)\star \frac{\partial (Q_{eff})_W(x,p)}{\partial \omega}\star (G_{eff})_W(x,p)\Big]
\label{NetaH2}
\end{align}
as already expressed in Eq. (\ref{generaltopologicalterm}). We will now proceed to show that Eq. (\ref{freetheoryrelations2}) indeed requires full Galilean invariance to hold and that consequently $\mathcal{N}_{\eta_H}$ as expressed in Eq. (\ref{NetaH1}) coincides with Eq. (\ref{NetaH2}) only in this case, while the more general case without Galilean boost invariance, implying a general functional dependence of the fermion bilinear operator Weyl symbol $(Q_{eff})_W$ on the kinetic energy operator, leads to coincidence modulo corrections suppressed by the inverse filling factor $\nu^{-1}$. Galilean invariance forces the dependence of the fermion bilinear operator Weyl symbol $(Q_{eff})_W$ on $S$ to be linear.\par
Note that, employing the Moyal star product in the form
\begin{align}
\star = \exp\Big(\frac{i}{2}B\epsilon_{ij}\overset{\leftarrow}{\partial}_{(D_W)_i}\overset{\rightarrow}{\partial}_{(D_W)_j}\Big)
\end{align}
we find the following commutation relation for covariant derivatives
\begin{align}
\nonumber &[(D_W)_i,(D_W)_j\delta^{jk}(D_W)_k]_{\star}=2iB\epsilon_{ij}(D_W)_j\\
&\Leftrightarrow\,\,(D_W)_j\delta^{jk}(D_W)_k\star (D_W)_i=(D_W)_l(\star (D_W)_j\delta^{jk}(D_W)_k\delta^{li}+2iB\epsilon_{li}).
\end{align}
The star subscript indicates that the Weyl symbols in the commutator are subjected to the Moyal star product. Rotational invariance implies that $(Q_{eff})_W$ and its inverse are in general functions of $(\omega ,S)$ where $\omega$ is frequency and $S=(D_W)_i\star g^{ij}\star (D_W)_j$. Consider analytical dependence of $(Q_{eff})_W$ on $S$. By linearity we may further focus on a monomial of $S$ of order $n$. Then we obtain
\begin{align}
\Big(\frac{\partial}{\partial g^{ij}}((D_W)_k\star g^{kl} (D_W)_l)^n\Big)\Big\rvert_{g^{ij}=\delta^{ij}}=\sum_{p=0}^{n-1}((D_W)_k\delta^{kl}(D_W)_l\star )^p(D_W)_{(i}\star (D_W)_{j)}(\star (D_W)_m\delta^{mn}(D_W)_n)^{n-1-p}
\end{align}
on the one hand and
\begin{align}
\nonumber &\frac{i}{2}\frac{\partial}{\partial p_{(i}}\Big(((D_W)_k\star g^{kl}(D_W)_l)^n\Big)\Big\rvert_{g^{ij}=\delta^{ij}}\star (D_W)_{j)}=\frac{1}{2}\frac{\partial}{\partial (D_W)_{(i}}\Big(((D_W)_k\star g^{kl}(D_W)_l)^n\Big)\Big\rvert_{g^{ij}=\delta^{ij}}\star (D_W)_{j)}\\
\nonumber &=\sum_{p=0}^{n-1}((D_W)_k\delta^{kl}(D_W)_l\star )^p(D_W)_{(i}(\star (D_W)_m\delta^{mn}(D_W)_n)^{n-1-p}\star (D_W)_{j)}\\
\nonumber &=\sum_{p=0}^{n-1}((D_W)_k\delta^{kl}(D_W)_l\star )^p(D_W)_{(i}\star (D_W)_{s_1}\Pi_{r=1}^{n-1-p}\Big(\star (D_W)_m\delta^{mn}(D_W)_n\delta^{s_rs_{r+1}}+2iB\epsilon^{s_rs_{r+1}}\Big)\delta^{s_{n-p}}_{j)}\\
&=\sum_{p=0}^{n-1}((D_W)_k\delta^{kl}(D_W)_l\star )^p(D_W)_{(i}\star (D_W)_{j)}(\star (D_W)_m\delta^{mn}(D_W)_n)^{n-1-p}+O\Big((D_W)^{2n-2}\Big)
\end{align}
on the other hand. Round brackets denote normalized symmetrization of uncontracted indices. In the simplest non-trivial case $n=2$ this amounts to
\begin{align}
\nonumber &\Big[\Big(\frac{\partial}{\partial g^{ij}}((D_W)_k\star g^{kl}(D_W)_l)^2\Big)-\frac{i}{2}\Big(\frac{\partial}{\partial p_{(i}}((D_W)_k\star g^{kl}(D_W)_l)^2\Big)\star (D_W)_{j)}\Big]\Big\rvert_{g^{ij}=\delta^{ij}}\\
&=2iB(D_W)_{(i}\star (D_W)_k\epsilon^{kl}\delta^{l}_{j)}.
\end{align} 
The relations in Eq. (\ref{freetheoryrelations2}) are therefore valid modulo commutators of covariant derivatives in the absence of Galilean boost symmetry. Consider now two functions $F(S)$ and $G(S)$. These fulfill
\begin{align}
F\star G-F\cdot G=\frac{i}{2}BF\overset{\leftarrow}{\partial}_{(D_W)_i}\overset{\rightarrow}{\partial}_{(D_W)_j}G+...
\end{align}
where dots mark higher order terms in the non-commutative gradient expansion. If the particle system under consideration reduces to a Fermi liquid in the limit $B\to 0$ with Fermi momentum $k_F\propto \sqrt{n}$ and particle number density $n$ we finally arrive at $F\star G-F\cdot G=O(Bk_F^{-2})=O(\frac{1}{\nu})$ for $F,G\sim O(1)$. The filling factor is given by $\nu =2\pi nl_B^2$ with magnetic length $l_B\sim 1/\sqrt{B}$. Applying our findings to the two expressions in Eqs. (\ref{NetaH1}) and (\ref{NetaH2}) representing $\mathcal{N}_{\eta_H}$ proves our claim. Note that the star product commutator $[F,G]_{\star}=O(\frac{1}{\nu})$ for $F,G\sim O(1)$ which will be relevant for the subsequent Appendix \ref{appendixE}.

\section{On the topological robustness of $\mathcal{N}_{\eta_H}$}
\label{appendixE}

In the presence of Galilean invariance and to leading order in filling fraction in the absence of Galilean boost symmetry we found in Appendix \ref{appendixD} that the Hall viscosity for a non-interacting integer quantum Hall system may be expressed as $\eta_H=\frac{1}{2\pi}\mathcal{N}_{\eta_H}B$ with coefficient
\begin{align}
\nonumber \mathcal{N}_{\eta_H}=&\frac{1}{16\pi^2\beta A}\int d^3x\int d^3p\frac{1}{B_{eff}}\epsilon^{ik}\delta^{jl}\Tr\Big[\frac{\partial (Q_{eff})_W(x,p)}{\partial p_i}\star (D_W)_j\star (G_{eff})_W(x,p)\\
&\star\frac{\partial (Q_{eff})_W(x,p)}{\partial p_k}\star (D_W)_l\star(G_{eff})_W(x,p)\star \frac{\partial (Q_{eff})_W(x,p)}{\partial \omega}\star (G_{eff})_W(x,p)\Big].
\label{possibletopologicalterm}
\end{align}
Importantly, while rotational invariance implies that $(Q_{eff})_W$ and its inverse are in general functions of $(\omega ,S)$ where $\omega$ is frequency and $S=(D_W)_i\star g^{ij}\star (D_W)_j$, Galilean invariance forces the dependence on $S$ to be linear.\par
We will now perform a variation of Eq. (\ref{possibletopologicalterm}) that keeps rotational and translational invariance intact modulo gauge field terms giving rise to a constant gauge curvature field. Since we know from explicit calculation that the expression for $\mathcal{N}_{\eta_H}$ is invariant under changes of the magnetic field $B$ which keep the chemical potential in the gap, we may restrict ourselves to variations at constant magnetic field. Covariant derivatives may then only change via infinitesimal local gauge transformations which is formally expressed as the contraint $\delta [(D_W)_i,(D_W)_j]_{\star}=0$ with the subscript $\star$ implying the commutator of star products of Weyl symbols. Since the expression for $\mathcal{N}_{\eta_H}$ in Eq. (\ref{possibletopologicalterm}) is gauge invariant, we may assume that the variation acts only on $(Q_{eff})_W$ and $(G_{eff})_W$ with $\delta (D_W)_i=0$. 
We introduce the short-hand notations (suppressing the subscript $_{eff}$ on the Weyl symbols)
\begin{align}
(d_W)_i=G_W\star\frac{\partial}{\partial (D_W)_i}Q_W,\,\,\,\,(d_W)_3=G_W\star\frac{\partial}{\partial \omega}Q_W,\,\,\,\,\Delta_W=G_W\star\delta Q_W
\end{align}
and perform the substitution $p_i\to D_i$ in Eq. (\ref{possibletopologicalterm}). Furthermore we will make use of the relations
\begin{align}
\delta G_W=-G_W\star \delta Q_W\star G_W
\end{align}
and
\begin{align}
\delta (d_W)_i=\frac{\partial}{\partial (D_W)_i}\Delta_W+[(d_W)_i,\Delta_W]_{\star}
\label{variationdW}
\end{align}
respectively. Denoting the integrand of Eq. (\ref{possibletopologicalterm}) after removing trivial factors by $I$, we find
\begin{align}
I=\epsilon^{ik}\delta^{jl}\Tr\Big[(d_W)_i\star (D_W)_j\star (d_W)_k\star (D_W)_l\star (d_W)_3\Big]
\end{align}
such that 
\begin{align}
\nonumber \delta I=&\epsilon^{ik}\delta^{jl}\Tr\Big[(\delta (d_W)_i)\star (D_W)_j\star (d_W)_k\star (D_W)_l\star (d_W)_3\\
\nonumber &+(d_W)_i\star (D_W)_j\star (\delta (d_W)_k)\star (D_W)_l\star (d_W)_3\\
&+(d_W)_i\star (D_W)_j\star (d_W)_k\star (D_W)_l\star (\delta (d_W)_3)\Big]\equiv\delta I_1+\delta I_2.
\end{align}
We introduce a splitting $\delta I=\delta I_1+\delta I_2$ corresponding to the two contributions in Eq. (\ref{variationdW}). After application of partial integration $\delta I_1$ may be written as
\begin{align}
\nonumber \delta I_1=&-\epsilon^{ik}\delta^{jl}\Tr\Big[\Delta_W\star \Big(\frac{\partial}{\partial (D_W)_i}((D_W)_j\star (d_W)_k\star (D_W)_l\star (d_W)_3)\\
\nonumber &+\frac{\partial}{\partial (D_W)_k}((D_W)_l\star (d_W)_3\star (d_W)_i\star (D_W)_j)\\
&+\frac{\partial}{\partial \omega}((d_W)_i\star (D_W)_j\star (d_W)_k\star (D_W)_l)\Big)\Big].
\end{align}
We may further decompose $\delta I_1=\delta I_1^1+\delta I_1^2$ with the former capturing the action of differentiation on explicit covariant derivatives while the latter contains the remainder. Employing the further relations
\begin{align}
\epsilon^{ij}\frac{\partial}{\partial (D_W)_i}(d_W)_j=-\epsilon^{ij}(d_W)_i\star (d_W)_j,\,\,\,\,\frac{\partial}{\partial (D_W)_i}(d_W)_3=-(d_W)_i\star (d_W)_3,\,\,\,\,\frac{\partial}{\partial \omega} (d_W)_i=-(d_W)_3\star (d_W)_i
\end{align}
we find after straightforward algebra
\begin{align}
\delta I_1^1=&\epsilon^{ij}\Tr\Big[\Delta_W\star\Big([(d_W)_i\star (D_W)_j,(d_W)_3]_{\star}-(D_W)_i\star [(d_W)_j,(d_W)_3]_{\star}\Big)\Big]\\
\nonumber \delta I_1^2=&\epsilon^{ik}\delta^{jl}\Tr\Big[\Delta_W\star\Big((D_W)_j\star (d_W)_i\star [(d_W)_k,(D_W)_l]_{\star}\star (d_W)_3\\
\nonumber &+[(D_W)_j,(d_W)_k]_{\star}\star (d_W)_3\star (d_W)_i\star (D_W)_l\\
&+[(D_W)_j,(d_W)_3\star (d_W)_k]_{\star}\star (d_W)_i\star (D_W)_l\Big)\Big]\\
\nonumber \delta I_2=&\epsilon^{ik}\delta^{jl}\Tr\Big[\Delta_W\star \Big([(D_W)_j,(d_W)_k\star (D_W)_l\star (d_W)_3\star (d_W)_i]_{\star}\\
&+[(D_W)_j,(d_W)_3\star (d_W)_i\star (D_W)_l\star (d_W)_k]_{\star}\Big)\Big].
\end{align}
According to our findings in Appendix \ref{appendixD} this result implies $\frac{\delta I}{I}=O(\frac{1}{\nu})$ and therefore that the leading term in filling fraction is topological under the assumed symmetries for the expression of $\mathcal{N}_{\eta_H}$ in Eq. (\ref{possibletopologicalterm}).

\end{document}